\documentclass[a4paper]{article}
\usepackage[table]{xcolor}
\usepackage[english]{babel}
\usepackage[utf8x]{inputenc}
\usepackage[T1]{fontenc}

\usepackage[a4paper,top=3cm,bottom=2cm,left=3cm,right=3cm,marginparwidth=1.75cm]{geometry}

\usepackage{amsmath}
\usepackage{setspace}
\usepackage{graphicx}
\usepackage[colorinlistoftodos]{todonotes}
\usepackage[colorlinks=true, allcolors=blue]{hyperref}
\usepackage{amssymb}
\usepackage{multirow}
\usepackage{dsfont}
\usepackage{amsthm}
\usepackage{siunitx}

\usepackage{imakeidx}

\makeindex

\usepackage{float}
\restylefloat{table}
\usepackage{makecell}

\usepackage{comment}

\usepackage{natbib}
\title{Tail-risk protection: Machine Learning meets modern Econometrics}

\author{
	Bruno Spilak 
	\footnote{IRTG 1792, School of Business and Economics, Humboldt-Universit{\"a}t zu Berlin, Dorotheenstr. 1, 10117 Berlin, Germany. Email: bruno.spilak@hu-berlin.de},
	Wolfgang Karl Härdle
	\footnote{corresponding author, W.I.S.E. - Wang Yanan Institute for Studies in Economics, Xiamen University, Xiamen, 361005, Fujian, China \newline
		C.A.S.E. - Center for applied Statistics and Economics, Humboldt-Universit{\"a}t zu Berlin, Unter den Linden 6, 10099 Berlin, Germany \newline
		Singapore Management University, 50 Stamford Road, 178899 Singapore, Singapore \newline
		Department of Mathematics and Physics, Charles University Prague, Ke Karlovu 2027/3, 12116 Praha 2, Czech \newline
		Email: haerdle@wiwi.hu-berlin.de 
	}
	\thanks{Financial support of the European Union's Horizon 2020 research and innovation program "FIN- TECH: A Financial supervision and Technology compliance training programme" under the grant agreement No 825215 (Topic: ICT-35-2018, Type of action: CSA), the European Cooperation in Science \& Technology COST Action grant CA19130 - Fintech and Artificial Intelligence in Finance - Towards a transparent financial industry, the Deutsche Forschungsgemeinschaft's IRTG 1792 grant, the Yushan Scholar Program of Taiwan and the Czech Science Foundation's grant no. 19-28231X / CAS: XDA 23020303 are greatly acknowledged. }
}

\begin{document}
\maketitle

\begin{abstract}

Tail risk protection is in the focus of the financial industry and requires solid mathematical and statistical tools, especially when a trading strategy is derived. Recent hype driven by machine learning (ML) mechanisms has raised the necessity to display and understand the functionality of ML tools. In this paper, we present a dynamic tail risk protection strategy that targets a maximum predefined level of risk measured by Value-At-Risk while controlling for participation in bull market regimes. We propose different weak classifiers, parametric and non-parametric, that estimate the exceedance probability of the risk level from which we derive trading signals in order to hedge tail events. We then compare the different approaches both with statistical and trading strategy performance, finally we propose an ensemble classifier that produces a meta tail risk protection strategy improving both generalization and trading performance.

\end{abstract}


\newpage

\section{Introduction}
\label{sec:intro}

"Black swan events" are occurring from time to time. After 2008, investors in the DAX had to wait 5 years before recovering their loss. The Nikkei stock index still has not yet recovered, even 25 years after the bubble burst. The cryptocurrency market has plunged drastically at the end of 2017. What choices are left to investors in these situations ?\\

Indeed, it is one of the main challenges of quantitative finance to build models and develop tools able to protect investments from such extreme events \cite{boll:2011}. The first risk management strategy addressing the latter is diversification but \citet{longin:2001} show that when financial markets exhibit huge downturn periods, correlation significantly increase, hence are counteracting the effect of portfolio diversification. Another strategy for tail risk protection is option based, where the risk manager is long in put options. Yet, loss aversion leads to high prices of put options (\citet{kozh:2013}, \citet{pack:2017}). A solution to escape this dilemma is to build a strategy with dynamic asset allocation. The risk manager creates an asymmetric risk profile with participation in upside market and protection against severe loss. This trading strategy is often called tail risk protection strategy by practitioners and is much more cost efficient and flexible than the options strategy \cite{Franke:2019}. In order to pursue such a strategy, one needs to predict distributional properties of the portfolio. This, of course, is obviously impossible since not only distributions but also their parameters change in time. This motivates our paper where we hint on comparing Econometrics tools, Machine learning (ML)\index{machine learning (ML)} and modern Local Parametric Approach (LPA)\index{local parametric approach (LPA)}. We present a dynamic tail risk protection strategy that targets a maximum predefined level of risk measured by Value-At-Risk while controlling for participation in bull market regimes. We propose different weak classifiers, parametric, based on GARCH tool, and non-parametric using ML, that estimate the exceedance probability of the risk level from which we derive trading signals in order to hedge tail events.\\

The GARCH\index{GARCH} tool with normally distributed innovations allows us to catch volatility clusters and yields good volatility forecasts. The application of Extreme Value Theory (EVT)\index{extreme value theory (EVT)} to GARCH residuals provides insight into the tail probabilities, that is, the likeliness of an extreme loss. Nevertheless, econometric models rely on strong assumptions and cannot deal with structural breaks that happen in financial time series.\\

ML has been successful in many applications during the last years thanks to its generalization power on large datasets, but, in quantitative finance, in particular for financial returns forecasting, ML did not prove its superiority in comparison with its application to text analysis or image recognition, where Deep Learning (DL)\index{deep learning (DL)} tools became state-of-the-art. Moreover, the black box effect of DL makes the industry reluctant to invest in such models. In theory, DL can extract information from non-linear relations in high dimensional space, so for quantitative finance practitioners, it should be natural to use the machine learning tool box. Nevertheless, DL models suffer from overfitting and can be very difficult to train. If they offer similar prediction accuracy than econometrics, it might be preferable to use the latter with regards to its interpretability, lower complexity and cheaper computational costs.\\

The cryptocurrency market has experienced an exponential growth during 2017, where BTC peaked at \$19 783.06 on December 17th 2017 and dropped below \$14 000 24 hours later, losing one third of its value. One year later, on December 7th 2018, BTC price briefly dipped below \$3 300, a 76\% drop from the previous year and a 15-month low. At the time of writing, the cryptocurrency market is experiencing a new large uptrend as swathes of institutional investors are gaining new interest because of the launch of new futures contracts on both regulated and unregulated cryptocurrency exchanges. Central banks also reacted to private initiatives of launching digital currencies, such as Facebook's Libra, as they may have the potential to dilute the main power of central banks - to control the supply of money to the economy. With high volatility and new interest in the cryptocurrency market, the BTC ecosystem is a perfect environment to show the effectiveness of tail risk strategy. Indeed, its high volatility promises high gains in upward movement of the price. Our goal is then to protect investors from large downturns of the market through an accurate prediction of tail risk.

Since Basel committee on Banking Supervision Amendment to incorporate market risk (1996), regulators imposed the use of certain metrics to measure the risk of investments, such as Value-At-Risk\index{Value-At-Risk (VaR)} ($\operatorname{VaR}$) and Expected-Shortfall  ($\operatorname{ES}$). In order to meet both investors' and regulators' will, we build a tail risk protection strategy that controls the $\operatorname{VaR}$ of our portfolio by ensuring that it is below a certain level, denoted as target $\operatorname{VaR}$.\\

As high volatility segments often precede market swings, the $\operatorname{GARCH}$\index{GARCH} model is a natural tool for tail risk protection, since $\operatorname{GARCH}$ catches these volatility clusters. Nevertheless, a volatility based risk management strategy forces us to divest in such period, reducing alpha possibilities in case of positive movement. Since financial returns have heavy tails, the $\operatorname{EVTGARCH}$\index{EVTGARCH} \cite{Mcneil:2000} allows us to improve our forecast of the tail event direction. Indeed, seeking alpha, excess return over a benchmark, has become more challenging for banks and institutional investors which follow strong regulations based on $\operatorname{VaR}$ estimates. We address this problem here by focusing on dynamic risk management based on econometrics, explaining stylized facts of financial time series with parametric approaches, and non- or local parametric methods, in particular the LPA from \citet{Spokoiny:2009a}, Multilayer Perceptron ($\operatorname{MLP}$) and Long Short-Term Memory ($\operatorname{LSTM}$) neural networks, which catches non-linear features with memory.\\

Our contribution is a rigorous comparison of $\operatorname{GARCH}$ and ML based tools in the context of extreme loss prediction based on their forecast in a classification scheme. On top, we show how to use standard Ensemble method as a meta classifier that produces a hybrid trail risk protection strategy, improving both generalization and trading performance by taking advantage of each approach. Finally, we evaluate our strategy with a realistic backtest including trading fees, by comparing it with classical buy-and-hold benchmark and other recent machine learning oriented tail risk protection strategies, such as the constant target VaR from \citet{rick:2019} and the Varspread strategy from \citet{pack:2017}.

The results showed here will certainly motivate practitioners to apply ML techniques in order to  improve $\operatorname{GARCH}$ performance. We provide comparison metrics such as forecast error, classification metrics and backtest results of our tail risk strategies on the cryptocurrency market with BTC investment. We also provide robustness checks through cross-validation. \\

This paper is structured as follows. First we review the current literature of our subject, then we explain the trading strategy we aim to build. In the third and fourth part, we explain the theoretical models used to build our strategy. In the final section, we present our results.

\section{Background \& Literature review}

\subsection{Volatility as a risk measure ? }\index{risk measure}

A good risk measure must be tailored to the investor's preferences\index{investor's preferences} which are often unknown in practice. Starting from preferences, practitioners often make assumptions in order to build an "optimal risk measure", corresponding to an imagined investor's goal. A large part of the financial literature studies volatility\index{volatility} as a risk measure, since it is nicely tied to Gaussian and LS techniques. For example, the VIX, referred to as the "investor fear gauge", is often taken as a sentiment indicator since volatility reflects investors' aversion to risk. The mean-variance portfolio, developed by Markowitz, is built under such assumptions where the weights of the risky assets included in the portfolio are derived from their volatility. Nevertheless, financial returns often have fat-tails and are not normally distributed, which Markowitz theory does not account for. Finally, volatility, being symmetric, is not realistic as a risk measure, since it does not take into consideration investors' loss aversion\index{risk aversion}, weighting equally volatility associated with gains and losses. Most investors are more concerned about downside risk, or losses, rather than volatility \cite{boll:2015}. In this paper, our goal is to build a trading strategy avoiding large losses or tail risk\index{tail risk} which is better suited for loss averse investors.

\citet{pack:2017} showed that a trading strategy accounting for tail risk\index{tail risk} can outperform simple buy-and-hold and traditional portfolio protection strategies. By using models such as GARCH with normally distributed innovations and GARCH with innovations following a Generalized Pareto Distribution (GPD), they built a new criterion for riskiness defined as the evolution of the estimated Value-At-Risk (VaR) spread between the two models. Thus, in period of increasing tail risk, this spread is significantly different from 0 which allows the trader to take adequate decisions. We denote this strategy as the Varspread strategy\index{Varspread strategy}. \citet{rick:2019} compares different risk measure such as volatility, VaR and Conditional-Value-at-Risk (CVaR), also named Expected Shortfall (ES), in order to build dynamic trading strategies and find that downside risk measures outperform volatility in terms of a higher Sharpe Ratio, better drawdown protection and higher utility gains for mean-variance and loss-averse investors. \citet{Happersberger:2019} also focus ES and VaR forecasts in order to manage dynamic tail risk protection strategies.

All the papers mentioned above and in general the literature of tail risk protection, focus on predicting risk measures in a regression manner and study the total distribution of the returns where the goal is to minimize Mean Squared Error (MSE) type measures of fit. Our argument is that it is not necessary to predict the total distribution of the return, whether it is crucial to correctly forecast the direction of the tail, in particular since in our case big profits or losses are at stake \cite{jorda:2011}.

\subsection{Investors' preferences}

\citet{rick:2019} developed a tail risk protection strategy, denoted target VaR strategy\index{target VaR strategy}, where the trading signals are calibrated so the VaR of the strategy is constant over time for a predefined significance level $\alpha$. The weights of the risky asset in the simple portfolio consisting of two assets, one risky and one riskyless asset, is a function of a constant level of risk defined as a VaR level denoted, target VaR, depending on investors' risk aversion measured by $\alpha$. The weights are calibrated so the VaR of portfolio is constant, equals to the target VaR. The target VaR strategy has the main advantage to be better interpretable for investors who can prescribe their acceptable loss limit to the trader. Nevertheless, it is assumed that investors' preferences are static since the target VaR is fixed and does not depend on $t$. Such strategy is based on the standard financial theory assumption that investors are rational and have invariant risk preferences.

However, with the development of Behavioral Finance, numerous studies draw attention on the fact that investors are often irrational and their preferences change with different situations. The early paper of \citet{kahn:1979} shows that investors are more risk averse with gains, but less with losses. More recently, \citet{wen:2014} study the characteristics of investors' risk preferences with different states of gains and losses and show that investors' risk preferences are time-varying with them. Indeed, the degree of risk aversion rises with the increasing gains and that of risk seeking improves with the increasing losses.

Taking into consideration investors' time-varying preferences\index{investor's preferences}, a constant target VaR strategy is inadequate. Indeed, we should decrease our target VaR in period of gains and increase it in periods of losses to allow more conservative trades in period of gains and more aggressive ones in case of losses.

\subsection{Dynamic tail risk protection strategy}

Our goal is to build a dynamic tail risk protection with an adaptive predefined level or risk, adapting to the benchmark's performance which influences investors' preferences, \citet{wen:2014}.  At constant significance level $\alpha$, when the benchmark is experiencing a period of good performance where the gains are increasing, its VaR is decreasing, thus the investors give more attention to smaller losses and tend to be more risk averse. As follows, the trader should adapt its strategy and aim for a smaller target VaR. In the inverse situation, during a period of losses, the tail risk of the benchmark is increasing and the investor only pay attention to large losses, tending to seek more risk. For example, one could use a threshold as a function of volatility in order to build a dynamic threshold labeling function as in \citet{Prado:2018}, chapter 3.3. In the next section, we will explain how to build such strategy.

\section{Trading strategy}
\label{sec:trading_strategy}
\subsection{Tail risk as a risk measure}

Let us give the following \textbf{tail risk definition}\index{tail risk}. Throughout this paper, we consider a risky asset, for example one stock, with price process $p_t$, where $t \in [0, ..., T]$ with $T$ the final period time step. As usual, we define the one period log-return, where one period is here one hour, as $r_t = \log \frac{P_t}{P_{t-1}}$ and the loss series $-r_t$.
As in \citet{pack:2017}, we define the Value-At-Risk for a \emph{risk level} $\alpha$, denoted $\operatorname{VaR}_t^\alpha$, as the $\alpha$-quantile of the distribution of the loss $-r_t$:

$$\mathrm{VaR}_{t}^{\alpha}=\inf \left\{l \in \mathbb{R}: \operatorname{P}\left(-r_{t}>l\right) \leq 1-\alpha\right\}=\inf \left\{l \in \mathbb{R}: \operatorname{P}\left(-r_{t} \leq l\right) \geq \alpha\right\}$$

If $r_t$ follows an absolutely continuous loss distribution, then the Value-At-Risk\index{Value-At-Risk} can be defined as an unlikely and severe loss which satisfies:

$$\operatorname{P}(- r_t \leq \mathrm{VaR}_{t}^{\alpha} ) = 1 - \alpha$$

where $\alpha$ is small corresponding to investors' risk aversion\index{risk aversion}. In practice, the risk level often takes the value of 0.01, 0.025 or 0.05. The Value-At-Risk characterizes the far right tail of the distribution of the loss, thus we use it as a measure of tail risk\index{tail risk}. For example, a $ \operatorname{VaR} = u_t^{\alpha} = 0.05$ for $ \alpha=0.01$ means that returns below -5\% only happen 1\% of the time. Moreover, for any threshold $u$, $\operatorname{P}(-r_t \geq u)$ is called the exceedance probability over $u$\index{exceedance probability}.

\subsection{Tail risk protection strategy}\index{tail risk protection strategy}

Now, let us consider a simple buy/sell trading strategy where the trader decides at time $t$ either to enter the market with a full position (he invests the totality of his available capital at time $t$ into BTC) or to stay out of the market (he sells all available BTC or does not invest capital at time $t$ in order to have all his capital in a risk free asset) based on the information available at time $t$ denoted $\mathcal{F}_{t}$. The trader's decision can be represented as a binary variable, or trading signal, $s_t \in \{0, 1\}$ where 1 corresponds to the decision to stay out of the market and 0 to the decision to fully invest the capital at time $t$. The return of such strategy is defined for all $t \in [1, \ldots, T]$ as $ R_t = (1 - s_{t-1})r_t$ and the excess return $R_t - r_t$.

\paragraph{Target VaR}\index{target VaR}
The goal of tail risk protection is to maximize the expected economic utility of a risk averse investor which can be represented by the risk-adjusted return of the strategy, where risk is characterized by the tail behavior of the portfolio. In other words, our goal is to lower the probability of tail risk\index{tail risk}, that is, at constant level $\alpha$, to have a lower portfolio VaR than the buy-and-hold benchmark strategy which buys the asset at the beginning of the period and sell it at time $T$, the end of the period of investment. In parallel, we have two choices to maximize the return, either we maximize the expected return, following the portfolio selection criterion from \citet{Markowitz:1952}, or we can maximize the total return at $T$, as suggested by \citet{Kelly:1956}. Since in our setting, we are optimizing the strategy for multiple periods \cite{Hakansson:1971} we choose the latter. We can write the following optimization program:
\begin{equation}
\begin{aligned}
\max & \quad  \sum_{t=1}^{T}(R_t - r_t) \\
\textrm{s.t.} & \operatorname{VaR}_t^\alpha \leq \operatorname{TVaR}_{t}^\alpha\\
\end{aligned}
\label{eq:or_opt_prg}
\end{equation}
where $\operatorname{VaR}^\alpha$ is the portfolio VaR and $ \operatorname{TVaR}_{t}^\alpha$ is the \textbf{target VaR for level} \pmb{$\alpha$} given by the investor.

We have the following equivalence,  $\forall t \in [1, \ldots, T]$:
\begin{align}
& \operatorname{VaR}_t^\alpha \leq \operatorname{TVaR}_t^{\alpha} \nonumber \\
& \operatorname{P}(- R_t \leq \operatorname{TVaR}_t^{\alpha}) \geq 1 - \alpha \label{eq:constraint}
\end{align}
Thus, one relaxes the constant constraint from \citet{rick:2019} and aims at constructing a trading mechanism with signal $s = \{s_1, \ldots, s_{T}\}$ that has a maximum $\operatorname{TVaR}^{\alpha}$. From now on, we think of fixed $\alpha$ and we will write $\operatorname{TVaR}_t^{\alpha}$ as  $\operatorname{TVaR}_{t}$ for the benchmark Value-At-Risk and  $\operatorname{VaR}_t^{\alpha}$ as  $\operatorname{VaR}_t$ for the portfolio.

\paragraph{How to define the signals $s_t$}
For all $t \in [0, T-1]$, we define:

\begin{equation}
s_{t} = \operatorname{\mathbf{I}}_{- r_{t+1} \geq \operatorname{TVaR}_{t+1} }
\label{def:signal}
\end{equation}

and by construction $R_{t} = (1 - \operatorname{\mathbf{I}}_{- r_t \geq  \operatorname{TVaR}_{t} }) r_t$. Since  $\operatorname{TVaR}_{t}$ is strictly positive for financial assets loss series, we have $ \operatorname{P}( - R_{t} \leq \operatorname{TVaR}_{t}  ) = 1$, thus \eqref{eq:constraint} is verified.

Since we do not know the true distribution of $r_t$ and, obviously, at time $t$, we do not observe $r_{t+1}$, we do not know $s_{t}$ and we must build estimates based on the observation available in order to make a trading decision. We estimate the following conditional probability, $\operatorname{P}( -r_{t+1} \geq \operatorname{TVaR}_{t+1} |  \mathcal{F}_{t})$, which is the exceedance probability\index{exceedance probability} over the threshold $\operatorname{TVaR}_{t+1}$, and decide whether the trader must close his position $s_t = 1$ or stay in the market $s_t = 0$.

\paragraph{Tail risk target $\operatorname{TVaR}^{\alpha}$}
The tail risk target, $\operatorname{TVaR}^{\alpha}$, is defined by the investors' preferences\index{investor's preferences}. Here, our goal is not to study investors' preferences, but to show how to build trading signals based on estimates of the exceedance probability\index{exceedance probability} over a given risk level. This problem has been well studied in the literature, for various applications, such as seismic risk assessment \cite{Honegger:2013}, risk assessment for decision making with application to terrorism \cite{Kunreuther:2002} or again floods, earthquakes, drought and hurricanes risk management \cite{Lambert:1994}, \cite{Mason:2007}. For trading strategies, \citet{Christoffersen:2006} and \citet{Linton:2007} considered exceedances above 0, whereas \cite{Chung:2007} focused on non-zero thresholds. In particular, \citet{Taylor:2016} studied thresholds that are not close to 0, as this is of greater relevance for risk management. This is also our interest but we consider extreme thresholds which are time varying and can be used as tail risk measure.

For illustration, we focus in this paper on different tail risk targets derived from the data itself. That is we use sample quantiles for different level $\alpha$ computed on the historical losses in a rolling window manner for different window size. In other words, to describe the investors' preferences for the next trading period $t+1$, we use the historical $\operatorname{VaR}$ of the loss series as tail risk measure defined for $w \in [1, T]$ as:

\begin{equation}
\forall t \in [w, T], \quad \forall i \in [t-w+1, t],   \operatorname{TVaR}_{t+1} = \operatorname{hist-\mathrm{VaR}}_{t}^{\alpha, w}=\inf \left\{l \in \mathbb{R}: \operatorname{P}\left(-r_{i}>l\right) \leq 1-\alpha\right\}
\label{eq:var_target_def}
\end{equation}

\begin{figure}[h!]
	\includegraphics[scale=0.4]{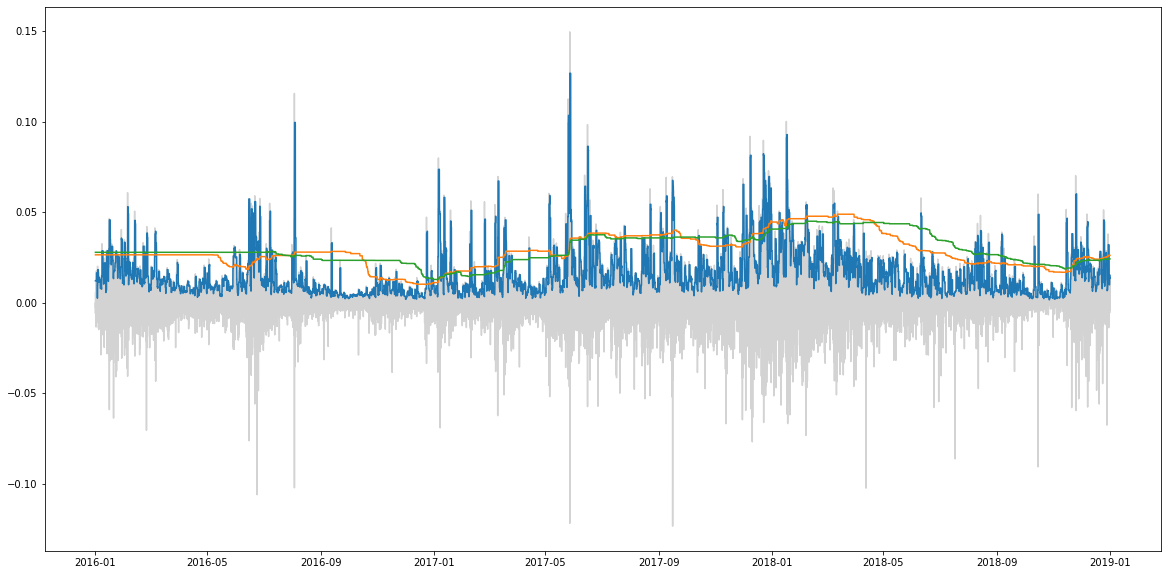}
	\centering
	\caption{\textcolor{gray}{Hourly btc losses} and $TVaR_t^{0.01, w}$ for different window size, \textcolor{blue}{$w = 24$} (one day), \textcolor{orange}{$ w = 2880$} (four months), \textcolor{green}{$w = 4320$} (six months) \protect \includegraphics[height=0.5cm]{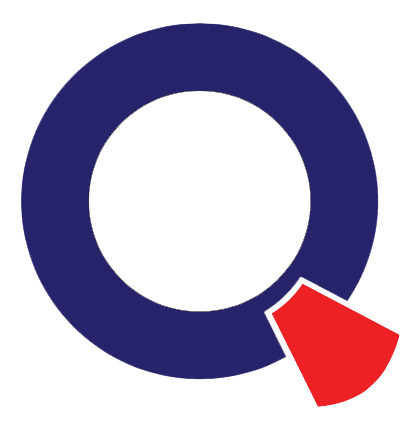} {\color{blue}\href{https://github.com/QuantLet/MLvsGARCH/}{MLvsGARCH}}}
	\label{fig:hist_var_window}
\end{figure}

The main advantage of this risk target is its computational simplicity allowing us to easily build training labels for our classifiers that we will develop in the next sections. On top, it is adaptable to investors' preferences, since using a small rolling window would suit an investor with varying preferences where the risk target quickly adapts itself to changes in the true return distribution, while large rolling windows give more stable risk target, corresponding to an investor with static preferences. In this paper, we will use three window sizes: 24 (one day), 2880 (four months) and 4320 (six months), since we are using hourly data, which are plotted on Figure \ref{fig:hist_var_window} for $\alpha = 0.01$. We give more details about the dataset in the next section. Our choice is motivated by the goal to study how the tools used here react to different levels of noise in the target variable.

As we can see on Figure \ref{fig:hist_var_window} the $\operatorname{TVaR}_t^{0.01, 24}$ is much more conservative than the $\operatorname{TVaR}_t^{0.01, w}$ with larger $w$, with much more losses exceeding the risk target. Indeed, we have respectively 0.045, 0.012 and 0.011 exceedance for the $\operatorname{TVaR}_t^{0.01, w}$ with w is 24, 2880 and 4320, where the exceedance is defined as $N_e = \frac{1}{T}\sum_{t = 1}^{T} - r_t \geq \operatorname{TVaR}_t^{\alpha}$, that is the proportion of data exceeding the threshold. A strategy with $\operatorname{TVaR}_t^{0.01, 24}$ as tail risk target should be much more conservative and hedges much more losses than the two other risk targets.

\paragraph{Trader's decision}
Finally, how to decide whether we should hedge our position for the next trading period ? We could build an estimation of the return at the next period, $r_{t+1}$, in a regression manner, nevertheless, we are only interested in the distribution of the tail. Thus, in this paper, we directly approximate the conditional probability classes\index{probability class} of the variable $s_{t}$, that is $p_{t}$, defined $\forall t \in [0, \ldots, T-1]$ as:
\begin{equation}
p_{t} = \operatorname{P}(s_{t} = 1 |  \mathcal{F}_{t}) = \operatorname{P}(- r_{t+1} \geq \operatorname{TVaR}_{t+1}^{\alpha}|  \mathcal{F}_{t})
\end{equation}
From $p_{t}$, we can make the trading decision $\hat{s}_{t}$ defined for a threshold $u \in \mathbb{R}$:
\begin{equation}
\hat{s}_{t}= \begin{cases} 
1, \ \text{if } p_{t}\geq u\\
0,  \ \text{otherwise}
\end{cases}
\label{eq:decision}
\end{equation}
We can now write the realized return of our strategy, for all $t \in [1, T]$:
\begin{equation}
	R_t = (1 - \hat{s}_{t-1})r_t
\end{equation}

\section{Machine learning trader}\label{sec:ml_trader}

Let $X_t \in \mathbb{R}^p$ be some input feature vector of dimension $p$ which summarize $\mathcal{F}_{t}$. The aim is to approximate the true unknown conditional probability $\operatorname{P}(s_t | X_t)$ with a learning algorithm $\mathcal{M}$ for different risk targets. We consider four different $\alpha$: 1\%, 2.5\%, 5\% and 10\%, and three rolling windows to compute $\operatorname{TVAR}$: 24, 2880, and 4320. In the next two sections, we present different models $\mathcal{M}$ we use as estimator for each risk target. We first present the dataset we use and then we explain the different models considered.

\subsection{Data}

We apply the proposed strategy on BTC\index{BTC}. We collected 36193 close prices from Poloniex exchange, using its API, from 2016-01-01 to 2020-02-16 on a hourly basis. We look at intraday frequency, since it is not rare to observe severe loss at such frequency, due to a relatively higher volatility in the cryptocurrency market compared to traditional assets.

We split the dataset into two sets, train, from 2016-01-01 to 2019-01-01 00:00:00, and test sets from 2019-01-01 01:00:00 to 2020-02-16 22:00:00. We keep a relatively large test set in order to produce a final robust estimation of the out-of-sample trading strategy performance with a large backtesting period of 9486 observations.

\subsection{Neural networks}\index{neural network}
\label{sec:nn}
We first use two different neural network architectures for the trading signals prediction, the Multi-Layer Perceptron (MLP)\index{multilayer perceptron (MLP)} and the Long short-term memory neural network (LSTM)\index{long short-term memory (LSTM)} architecture from \citet{hoch:1997}. Neural networks are non-parametric, which allows us to avoid making strong assumptions on the data that are not met in reality and that is often the case for financial time series, for example with the normality assumption of financial returns. On top, neural networks are universal approximators \cite{Leshno:1993} which means that, in theory, neural networks can arbitrarily closely approximate the true process of $r_t$. In particular, LSTM neural networks are state-of-the-art for many applications such as speech recognition, text extraction, translation or handwriting recognition, since plain recurrent neural networks are not suited to non-stationary time series modeling. In finance, Deep learning\index{deep learning (DL)} has been used in various research, in particular by \citet{Franke:1999} and \citet{Zhang:2020} for portfolio management or by \cite{Kim:2018} for volatility forecasting with LSTM in comparison with GARCH. Nevertheless, one has yet to prove the superiority of neural networks compared to simpler parametric models in their application.

\subsubsection{Trader's function mapping}
\paragraph{Output}
In order to train the neural networks, we need to build the training input-output pairs. We could directly use the trading signal $s_t$ from \eqref{def:signal} as output labels, nevertheless, since $\operatorname{TVaR}_t^{\alpha, w}$ belongs to the tail of the loss series, $s_t$ suffers from sever class imbalance since $\frac{\sum_{t = w+1}^{T} \operatorname{\mathbf{I}}_{-r_t > \operatorname{TVaR}_t^{\alpha, w}}}{T} \ll	 0.5$, which is difficult to handle for machine learning models \cite{Krawczyk:2016}. On top, financial returns follow an asymmetric distribution and suffers from the leverage effect \cite{Black:1976}, \cite{Christie:1982}, commonly defined as volatility rising more rapidly when returns are negative than positive. To address that problem, we introduce a new category for $r_t$, when $r_t$ belongs to the right tail of the distribution, which allow us to control the misclassification cost of the final model. Indeed, let us introduce the following output variable defined as:
\begin{equation}
Y_{t}= \begin{cases} 
1, \ \text{if } - r_{t} < \operatorname{TVaR}_t^{1-\alpha, w}\\
2, \ \text{if } - r_{t} > \operatorname{TVaR}_{t}^{\alpha, w}\\
0,  \ \text{otherwise}
\label{eq:label_def}
\end{cases}
\end{equation}
We can define the predicted strategy signals as a function of the predicted output variable $\hat{Y}$ as follows :
\begin{equation}
\hat{s}_{t}= \begin{cases} 
1, \ \text{if } \hat{Y}_{t} = 2\\
0,  \ \text{otherwise}
\end{cases}
\end{equation}
and we easily derive the unconditional misclassification costs\index{misclassification cost} matrix in terms of excess returns.

$$
\begin{array}{ccccc}
\hline
\hline
& & \multicolumn{3}{c}{ \text{ Prediction }} \\
& & \text { 0 } &\text { 1 }  &\text { 2 } \\

\multirow{2}{*}{ \text { Outcome }} & \text { 0 } & 0 & 0  & \bar{r}_0  \bar{p}_0\\
& \text { 1 } &  0 & 0  & \bar{r}_1 \bar{p}_1\\
& \text { 2 } &  0 & 0  & \bar{r}_2 \bar{p}_2\\
\hline
\hline
\end{array}
$$
\setstretch{1.25}\\\\

where $\bar{r}_i$ is the expected return on class $i$ estimated with $\frac{1}{\sum_{t=w+1}^{T} \operatorname{\mathbf{I}}_{Y_t = i}}\sum_{t=w+1}^{T}r_t \operatorname{\mathbf{I}}_{Y_t = i} $ and $\bar{p}_i$ is the weight of class $i$ defined as $\frac{\sum_{t=w+1}^{T} \operatorname{\mathbf{I}}_{Y_t = i}}{T - w}$ which is the empirical estimate of the unconditional probability $\operatorname{P}(Y_t = i)$.

\begin{table}[h!]
	\begin{center}
		\begin{tabular}{ccrrr}
			\hline
			\hline
			& & \multicolumn{3}{c}{ \text{ Prediction }} \\
			$\alpha$ & $w$ (hours) & 0 & 1 & 2 \\
			\hline
			
			\multirow{3}{*}{0.01}   & 24       & 0.01 & \textcolor{red}{0.15} &  \textcolor{blue}{-0.15} \\
			& 2880   & 0.01 & \textcolor{red}{0.14} &  \textcolor{blue}{-0.14} \\
			& 4320   & 0.01 & \textcolor{red}{0.13} &  \textcolor{blue}{-0.14} \\
			\\
			\multirow{3}{*}{0.025}   & 24       & 0.01 & \textcolor{red}{0.15} &  \textcolor{blue}{-0.15} \\
			& 2880   & 0.01 & \textcolor{red}{0.14} &  \textcolor{blue}{-0.14} \\
			& 4320   & 0.01 & \textcolor{red}{0.13} &  \textcolor{blue}{-0.14} \\
			\\
			\multirow{3}{*}{0.05}   & 24       & 0.01 & \textcolor{red}{0.15} &  \textcolor{blue}{-0.15}\\
			& 2880   & 0.01 & \textcolor{red}{0.14} &  \textcolor{blue}{-0.14}\\
			& 4320   & 0.01 & \textcolor{red}{0.13} &  \textcolor{blue}{-0.14}  \\
			\\
			\multirow{3}{*}{0.1}     & 24  & 0.01 & \textcolor{red}{0.15} &  \textcolor{blue}{-0.15} \\
			& 2880   & 0.01 & \textcolor{red}{0.14} & \textcolor{blue}{-0.14}\\
			& 4320   & 0.01 & \textcolor{red}{0.13} & \textcolor{blue}{-0.14}\\
			
			\hline
			\hline
		\end{tabular}	
	\end{center}
	\caption{Class 2 misclassification costs for different $\operatorname{TVaR}^{\alpha, w}$}
	\label{table:class_cost}
\end{table}

As expected, when using the risk target defined in \eqref{eq:var_target_def}, we can see on table \ref{table:class_cost} that one has negative costs for the correct classification of class 2 since $\bar{r}_2 \leq 0$ and by definition, the conditional cost for predicting class 2 instead of 1 is much higher than the one for predicting class 2 instead of 0, since in the former, we miss the opportunity to invest during an extremely positive return period. Thus, we clearly see that the classifier must correctly predict the sign of the tail events in order to maximize the strategy return. In order to do so, we use a 3 neurons output layer with softmax activation function corresponding to the output variable $Y_t$ defined in \eqref{eq:label_def}.

\paragraph{Input layer}

As we are interested in understanding whether neural networks can, by themselves, learn valuable features to produce trading signals and feature selection is not our interest in this work, we use simple transformations of the returns, which can be seen as momentum features.

In particular, since price time series are not stationary, we use multiple returns instead, $X_t^p = r_{t-p}$ for different periods $p$, where $r_{t-p} = p_{t}/p_{t-1-p}-1$ and $p \in \{0, 1, 2, 4, 6, 13\}$. We also use the normalized difference between the returns and the two class thresholds which should help identifying a risk-buildup situation, when the returns either explodes toward the upper class threshold $\operatorname{TVaR}_t^{1 - \alpha, w}$ or severely drops toward the lower threshold $\operatorname{TVaR}_{t}^{\alpha, w}$, which are defined respectively as $U_t^{\alpha, w} = \frac{r_{t} - \operatorname{TVaR}_{t}^{\alpha, w}}{\operatorname{TVaR}_{t}^{\alpha, w}}$ and $D_t^{\alpha, w} = \frac{r_{t} - \operatorname{TVaR}_t^{1-\alpha, w}}{\operatorname{TVaR}_t^{1-\alpha, w}}$.

For the MLP, we then use the vector $X_t \in \mathbb{R}^{8} = (X_t^0, X_t^1, X_t^2, X_t^4, X_t^6, X_t^{13}, D_t^{\alpha, w}, U_t^{\alpha, w})$ as input for each risk target. For the LSTM model, since it is a recurrent neural network, we take advantage of its ability to directly modelize sequential data and we use the same features as above on a certain historical window of length 24, using one historical day to make a prediction for the next hour. Thus the features become: $X_t^p = (r_{t-p-23}, \ldots, r_{t-p})$ for $p \in \{0, 1, 2, 4, 6, 13\}$ and $U_t^{\alpha, w} = (\frac{r_{t-23} - \operatorname{TVaR}_{t}^{\alpha, w}}{\operatorname{TVaR}_{t}^{\alpha, w}}, \ldots, \frac{r_{t} - \operatorname{TVaR}_{t}^{\alpha, w}}{\operatorname{TVaR}_{t}^{\alpha, w}})$ and $D_t^{\alpha, w} = (\frac{r_{t-23} - \operatorname{TVaR}_t^{1-\alpha, w}}{\operatorname{TVaR}_t^{1-\alpha}}, \ldots, \frac{r_{t} - \operatorname{TVaR}_t^{1-\alpha, w}}{\operatorname{TVaR}_t^{1-\alpha}})$. We use the same input vector as for the MLP, but now $X_t$ is in $\mathbb{R}^{24, 8}$. We explain the hidden layer architecture in the next section.

Finally, we train our classifier, using Keras python library,  on the training data $D = \{(X_{w + 1}, Y_{w + 1}), \ldots, (X_T, Y_T)\}$ with Adam algorithm, which is a stochastic gradient descent method based on adaptive estimation of first-order and second-order moments of the gradient \cite{Kingma:2015}  and 128 batch size.

\subsection{Model selection}

\subsubsection{Hidden layers architecture}

\begin{figure}[h!]
	\includegraphics[scale=0.3]{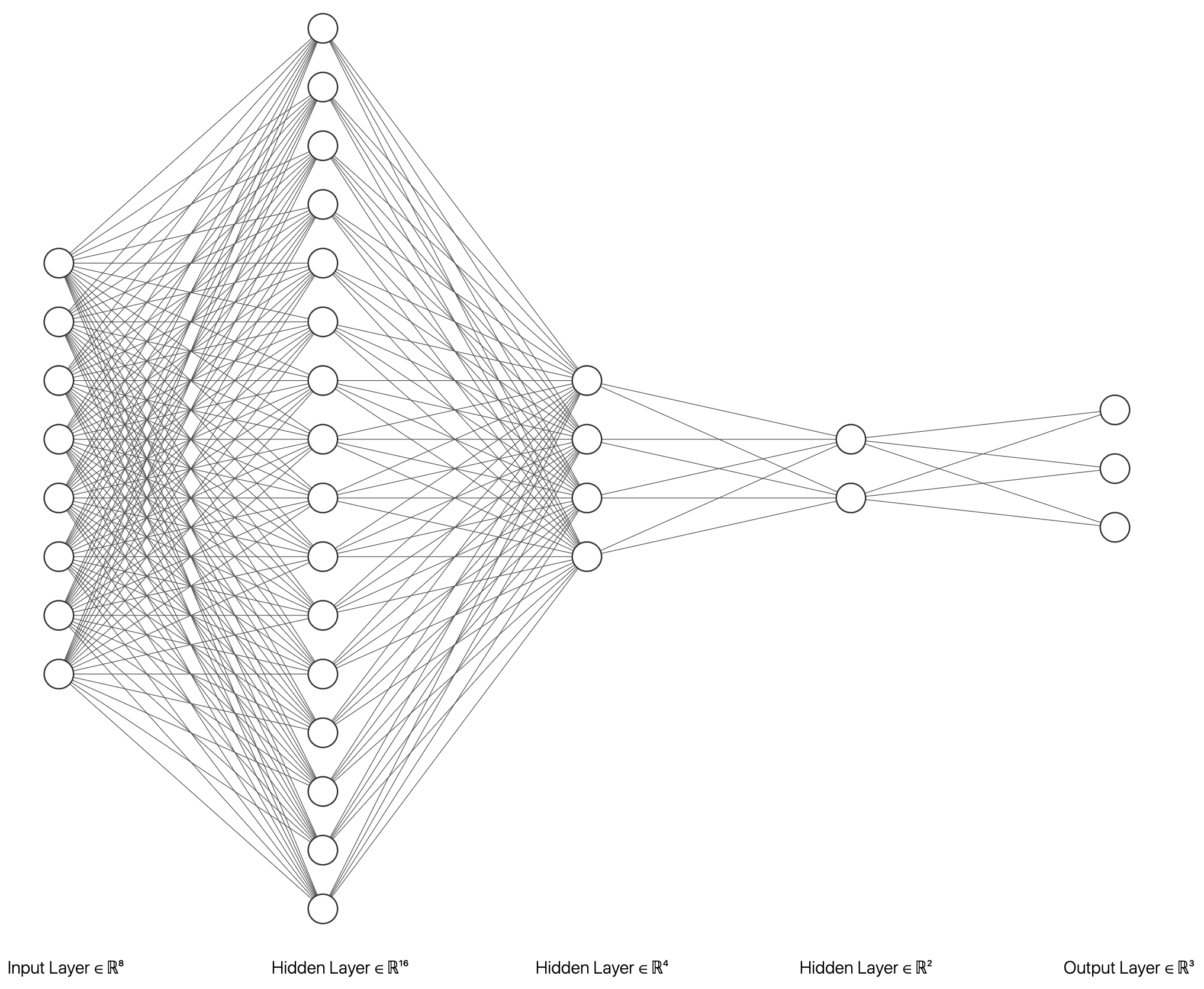}
	\centering
	\caption{Final MLP architecture}
	\label{fig:mlp_architecture}
\end{figure}

\begin{figure}[h!]
	\includegraphics[scale=0.5]{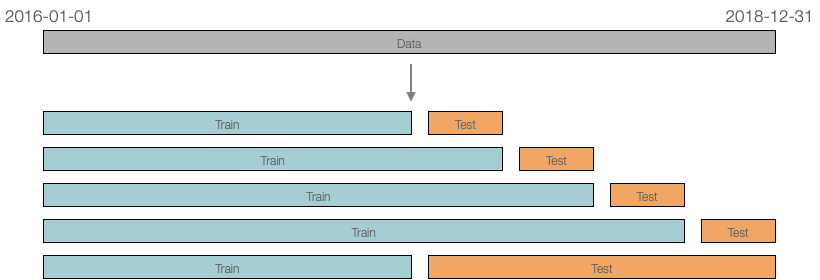}
	\centering
	\caption{Time series cross-validation with four folds}
	\label{fig:time_series_cv}
\end{figure}

On the train set we performed 10 folds cross-validation\index{cross-validation} for time series, described on figure \ref{fig:time_series_cv}, with one fold corresponding to one month, in order to test different hidden layers  architectures for the MLP model. In this work, we keep the hidden layers relatively simple in order to show how a single architecture can effectively extract features for different risk target output. Based on the cross-validation performance evaluated with AUC score (area under the ROC curve)\index{AUC}\index{ROC}, we choose the model with highest AUC score between class 1 and 2, while still controlling for robustness. That is we select the model with highest median AUC score between class 1 and class 2. After a small tuning, we keep 3 hidden layers, three fully connected layers with 16, 4 and 2 neurons respectively, as on the Figure \ref{fig:mlp_architecture}

As regularization technique, we use the state-of-the-art Dropout layer after the first and the second hidden layers with a dropout rate of 0.2, in order to prevent overfitting \cite{Srivastava:2014}.

As for the LSTM architectures, we simply change the first two fully connected layers to LSTM units in order to test whether LSTM can directly improve the performance of the MLP models. We use the hyperbolic tangent as activation function for the hidden layers.

\subsubsection{Strategy parameters and final decision}

Any classifier $\mathcal{M}$ trained on the dataset $D$ gives for output the probability class $p_t$\index{probability class}. Thus the trading strategy based on $\mathcal{M}$ has one parameter, the probability threshold $u$ from which we can make a trading decision with respect to \eqref{eq:decision}. The optimal parameter must solve the investor's goal define in \eqref{eq:or_opt_prg} with respect to their risk target. We can now reformulate the optimization problem as it follows:

\begin{equation}
\begin{aligned}
\max_{u, w} & \sum_{t=w+1}^{T}(R_t - r_t) =  \sum_{t=w+1}^{T}\{(1 - \hat{s}_t)r_t\}\\
\textrm{s.t.} & \operatorname{VaR}_t^\alpha \leq \operatorname{TVaR}_t^{\alpha}\\
\end{aligned}
\label{eq:opt_prg_param}
\end{equation}

\paragraph{Probability threshold $u$}
Denote $\mathcal{U}_{\alpha}$ the set of $u$ which are tail loss optimal, that is the set of $u$ respecting the constraint formulated in $\eqref{eq:constraint}$. Let us define our strategy with the trading signals $\{\hat{s}_{w+1}, \dots, \hat{s}_T\}$ and corresponding return $S_{w,\alpha} = \{R_{w+1}, \dots, R_T\}$ for a predefined risk target with level $\alpha$. We wish to define:
$$\mathcal{U}_{\alpha} =  \{u \in \mathbb{R},\quad \operatorname{P}( R_{t} \leq \operatorname{TVaR}_{t}^{\alpha}) \geq 1 - \alpha \}$$
Let us introduce the conditional probability classes for an arbitrary threshold $u \in \mathbb{R}$:\index{probability class}
\setstretch{2}
$$
\begin{array}{cccc}
\hline
\hline
& & \multicolumn{2}{c}{ \text{ Prediction }} \\
& & \text { 0/Hold } &\text { 1/Sell } \\

\multirow{2}{*}{ \text { Outcome }} & \text { 0/Hold } & \operatorname{TNR}(u)=\operatorname{P}(\hat{s}_t = 0 | s_t = 0) &  \operatorname{FPR}(u)=\operatorname{P}(\hat{s}_t = 1 | s_t = 0) \\
& \text { 1/Sell } &  \operatorname{FNR}(u)=\operatorname{P}(\hat{s}_t = 0 | s_t = 1) &  \operatorname{TPR}(u)=\operatorname{P}(\hat{s}_t = 1 | s_t = 1) \\ 
\hline
\hline
\end{array}
$$
\setstretch{1.25}

where $\operatorname{TNR}(u)$,  $\operatorname{FPR}(u)$,  $\operatorname{FNR}(u)$ and  $\operatorname{TPR}(u)$ stand for \textit{True Negative}, \textit{False Positive}, \textit{False Negative} and  \textit{True Positive Rates} for the threshold $u$ respectively. In statistical tests,  $\operatorname{FPR}(u)$ and $\operatorname{FNR}(u)$ are often called type I and type II errors respectively. The optimal threshold is given by the optimal trade-off between $\operatorname{FPR}$ and $\operatorname{FNR}$, but in order to respect the risk target, one cannot choose any trade-off, since, as one can easily see, $\operatorname{FNR}(u)$ corresponds to the exceedance probability of the portfolio over the risk limit. In practice, as it is pointed out in \cite{pack:2017}, such a strategy cannot avoid tail events that occur "totally out of the blue" (type II error in our formulation). Also, given the empirical stylized fact that return data feature little or no autocorrelation implies that such a strategy may signal a sell order, when ex-post holding the position would have been optimal (type I error).

We can easily see that any threshold $u^* \in \mathcal{U}_{\alpha}= \{u \in \mathbb{R},\quad \operatorname{TPR}_(u) \geq {}  \frac{\operatorname{P}(- r_t > \operatorname{TVaR}_{t}^{\alpha}) - \alpha}{\operatorname{P}(- r_t > \operatorname{TVaR}_{t}^{\alpha})} \}$ is tail loss optimal.\index{tail loss optimal}

\begin{proof}
	We need to compute the following probability for all $u \in \mathbb{R}$ and $t \in [w, T]$:
	\begin{align*}
	\operatorname{P}( R_{t} \leq\operatorname{TVaR}_{t}^{\alpha}) &  = 1 - \operatorname{P}( R_{t} >\operatorname{TVaR}_{t}^{\alpha})
	\end{align*}
	We have: 
	\begin{align}
	\nonumber \operatorname{P}( R_{t} >\operatorname{TVaR}_{t}^{\alpha}) & = \operatorname{P}( R_{t} >\operatorname{TVaR}_{t}^{\alpha} \cap \ s_{t}  = 1) +
	\nonumber \operatorname{P}( R_{t} >\operatorname{TVaR}_{t}^{\alpha} \cap \ s_{t}  = 0)\\
	\nonumber & = \operatorname{P}( R_{t} >\operatorname{TVaR}_{t}^{\alpha} | \ s_{t}  = 1) \operatorname{P}(s_{t}  = 1)  +
	\nonumber \underbrace{\operatorname{P}( R_{t} >\operatorname{TVaR}_{t}^{\alpha} | \ s_{t}  = 0)}_{= 0} \operatorname{P}(s_{t}  = 0) \\
	& = \operatorname{P}( R_{t} >\operatorname{TVaR}_{t}^{\alpha} | \ s_{t}  = 1) \underbrace{\operatorname{P}(s_{t}  = 1)}_{=\operatorname{P}(- r_t >\operatorname{TVaR}_{t}^{\alpha})} \label{esti_exceed_proba_eq}
	\end{align}
	and:
	\begin{align}
	\nonumber\operatorname{P}( R_{t} >\operatorname{TVaR}_{t}^{\alpha} | \ s_{t}  = 1)  = {} & \operatorname{P}( R_{t} >\operatorname{TVaR}_{t}^{\alpha} \cap \ \hat{s}_{t} = 0 | \ s_{t}  = 1) \\
	\nonumber&  + \operatorname{P}( R_{t} >\operatorname{TVaR}_{t}^{\alpha} \cap \ \hat{s}_{t} = 1 | \ s_{t}  = 1) \\
	\nonumber=  {} &\underbrace{\operatorname{P}( R_{t} >\operatorname{TVaR}_{t}^{\alpha} | \ \hat{s}_{t}  = 0 \cap s_{t}  = 1 ) }_\textrm{= 1}  \operatorname{P}(\hat{s}_{t}  = 0 | s_{t}  = 1) \\
	\nonumber& + \underbrace{\operatorname{P}( R_{t} >\operatorname{TVaR}_{t}^{\alpha} | \ \hat{s}_{t}  = 1 \cap s_{t}  = 1)}_\textrm{= 0}  \operatorname{P}(\hat{s}_{t}  = 1 | s_{t}  = 1 ) \\
	\nonumber= {} &   \operatorname{P}(\hat{s}_{t}  = 0 | s_{t}  = 1) \\
	\nonumber= {} &   \operatorname{FNR}(u) \\
	= {} &   1 - \operatorname{TPR}(u) \label{tpr_eq}
	\end{align}
	By plugging \eqref{tpr_eq} in \eqref{esti_exceed_proba_eq}, we have:
	\begin{align*}
	\nonumber \operatorname{P}( R_{t} >\operatorname{TVaR}_{t}^{\alpha}) & = \{ 1 - \operatorname{TPR}(u) \} \operatorname{P}(- r_t >\operatorname{TVaR}_{t}^{\alpha})
	\end{align*}
	Hence:
	\begin{align*}
	\operatorname{P}( R_{t} \leq\operatorname{TVaR}_{t}^{\alpha}) &  = 1 - \{ 1 - \operatorname{TPR}(u) \} \operatorname{P}(- r_t >\operatorname{TVaR}_{t}^{\alpha})\\
	& = 1 - \operatorname{P}(- r_t >\operatorname{TVaR}_{t}^{\alpha}) + \operatorname{P}(- r_t >\operatorname{TVaR}_{t}^{\alpha}) \operatorname{TPR}(u)
	\end{align*}
	In order to achieve the goal of our tail risk protection strategy, we must have: 
	\begin{align}
	\nonumber  & \quad \operatorname{P}(R_t \leq \operatorname{VaR}_t^{\alpha, target}) \geq {} 1 - \alpha \\
	\nonumber    &  \quad 1 - \operatorname{P}(- r_t >\operatorname{TVaR}_{t}^{\alpha}) + \operatorname{P}(- r_t >\operatorname{TVaR}_{t}^{\alpha}) \operatorname{TPR}(u) \geq {}  1 - \alpha \\
	& \quad   \operatorname{TPR}(u) \geq  \frac{\operatorname{P}(- r_t >\operatorname{TVaR}_{t}^{\alpha}) - \alpha}{\operatorname{P}(- r_t >\operatorname{TVaR}_{t}^{\alpha})}\label{eq:tpr_constraint}
	\end{align}
	Thus any threshold $u^*$ with satisfies \eqref{eq:tpr_constraint} is tail loss optimal, since it will ensure a maximum target $\operatorname{VaR}$ by controlling the type II error.
	
\end{proof}

Finally, to decide for the final probability class threshold $u^*$, we simply choose the threshold that maximize the cumulative excess return on the train set and apply it on the corresponding test set.

\begin{figure}[h!]
	\includegraphics[scale=0.5]{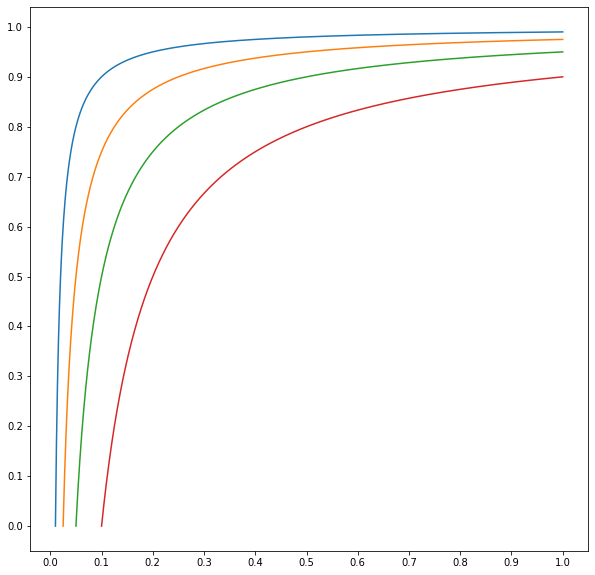}
	\centering
	\caption{Expected exceedance and corresponding minimum $\operatorname{TPR}$ for significance levels \textcolor{blue}{0.01}, \textcolor{orange}{0.025}, \textcolor{green}{0.05}, \textcolor{red}{0.1} \protect \includegraphics[height=0.5cm]{qletlogo_tr.png} {\color{blue}\href{https://github.com/QuantLet/MLvsGARCH/}{MLvsGARCH}}}
	\label{fig:min_tpr_exceedance}
\end{figure}

For illustration, on Figure $\ref{fig:min_tpr_exceedance}$, we plotted the shape of the constraint on $\operatorname{TPR}$ with respect to the risk limit of the investor for different risk level $\alpha$. As we can see, as the target risk for a level $\alpha$ diverge from the $\alpha$-quantile of the true distribution of the returns, the constraint on the classifier performance strengthens.

As described in section \ref{sec:trading_strategy}, the risk targets are defined by Equation $\eqref{eq:var_target_def}$. In the Table \ref{tbl:min_tpr_exceedance_alpha_window}, we show the different constraints from equation \eqref{eq:constraint} on the classifier in order to achieve the goal of the tail loss strategy based on the hourly returns of btc on the train set.

\begin{table}[H]
\begin{center}
	\begin{tabular}{cccc}
		\hline
		\hline
		$\alpha$ & $w$ (hours) & exceedance & min $\operatorname{TPR}$ (\%) \\
		\hline
		 \multirow{3}{*}{0.01} & 24 & 4.5 & 78 \\
		& 2880 & 1.2 & 18 \\
		& 4320 & 1.1 & 11 \\
		\multirow{3}{*}{0.025} & 24 & 5.5 & 55 \\
		& 2880 & 2.7 & 7 \\
		& 4320 & 2.6 & 5 \\
		\multirow{3}{*}{0.05} & 24 & 8.0 & 38 \\
		& 2880 & 5.2 & 4 \\
		& 4320 & 5.1 & 3 \\
		\multirow{3}{*}{0.1} & 24 & 12.5 & 20 \\
		& 2880 & 9.7 & 0 \\
		& 4320 & 9.4 & 0 \\
		\hline
		\hline
	\end{tabular}
	
\end{center}
\caption{Expected exceedance and corresponding minimum $\operatorname{TPR}$ for different $\operatorname{VaR}$ estimators with parameters $(\alpha, w)$, $w=2880$ and $w=4320$ correspond to 4 months and 6 months respectively \protect \includegraphics[height=0.5cm]{qletlogo_tr.png} {\color{blue}\href{https://github.com/QuantLet/MLvsGARCH/}{MLvsGARCH}}}
\label{tbl:min_tpr_exceedance_alpha_window}
\end{table}

As we can see, using a small rolling window to compute the risk target is more conservative thus the classifier must have a large $\operatorname{TPR}$ in order to ensure a maximum risk target of the final strategy.
\section{Auto-regressive models}

Since neural networks are complex non-linear multi-factor models, it is justified to ask if a simple parametric model calibrated on the same training set can catch tail-events as good ML methods or better. In order to compare both approaches, we used three different parametric and a LPA technique from \citet{Spokoiny:2009a}, that we present in this section.

\subsection{ARMA-GARCH type models}\index{ARMA-GARCH}\index{GARCH}

\subsubsection{ARMA-GARCH}
As a benchmark, we first used an $\operatorname{ARMA-GARCH}$ model for BTC loss series. Indeed, $\operatorname{GARCH}$, developed by \citet{Bollerslev:1986}, catches well volatility clusters, which often appears in returns time series and $\operatorname{ARMA}$ model allows for a time-varying mean of the returns. The $\operatorname{ARMA(P,Q)-GARCH(q,p) }$ model is defined as:
$$
\begin{aligned}
r_{t} &=a_{1} r_{t-1}+a_{2} r_{t-2}+\ldots+a_{P} r_{t-P} \\ &+\varepsilon_{t}+b_{1} \varepsilon_{t-1}+b_{2} \varepsilon_{t-2}+\ldots+b_{Q} \varepsilon_{t-Q} 
\end{aligned}$$
with
$$\begin{aligned}
\varepsilon_{t} &=Z_{t} \sigma_{t} \\ Z_{t} & \sim N(0,1) \\ \sigma_{t}^{2} &=\omega+\sum_{i=1}^{p} \beta_{i} \sigma_{t-i}^{2}+\sum_{j=1}^{q} \alpha_{j} \varepsilon_{t-j}^{2}
\end{aligned}$$
where $\omega > 0$, $\alpha_i \geq 0$, $\beta_i \geq 0$ and $\sum_{i=1}^{p} \beta_{i}+\sum_{j=1}^{q} \alpha_{j}<1$.

\subsubsection{ARMA-EVTGARCH}\index{ARMA-EVTGARCH}\index{EVTGARCH}

Assuming normally distributed innovations, $Z$, is often not justified and, since in our case we are interested in modeling extreme events, we also fitted an $\operatorname{ARMA-GARCH}$ model with a Generalised Pareto distribution (GPD)\index{generalised Pareto distribution} for the innovations of the $\operatorname{GARCH}$ process. This approach has been studied numerous time in the literature, in particular in \citet{pack:2017} or \citet{Taylor:2016} for risk management in trading. We refer to that model as $\operatorname{ARMA-EVTGARCH}$.

The distribution function of the GPD is given by:
$$G_{\xi, \beta}(x)=\begin{cases} 
1 - (1 + \xi x / \beta)^{-1/\xi}, \quad \xi \ne 0 \\
1 - \exp^{-x/\beta},\quad \xi = 0 
\end{cases}$$
where $\beta > 0$,$ x \geq 0$ when $x \geq 0$ and $0 < x \leq -\beta/\xi$ when $\xi < 0$.

The GPD describes the tail of the data since only the extremes outcomes are included in the estimation with the method of threshold exceedances as we explain in the next paragraph.

\subsubsection{Model selection, calibration and exceedance probability}

To find the proper lag orders, $P$ and $Q$, for the mean, and $p$ and $q$, for the volatility processes, we use the classical Box-Jenkins method and AIC criterion, see \citet{Chen:2016} for a detailed explanation. We perform the model selection on the last 2280 observations (four months) of the train set. The final model selected is an $\operatorname{ARMA(3, 1)-GARCH(1,2)}$, we use the same orders $\operatorname{ARMA-EVTGARCH}$.

\begin{figure}[h!]
	\includegraphics[scale=0.2]{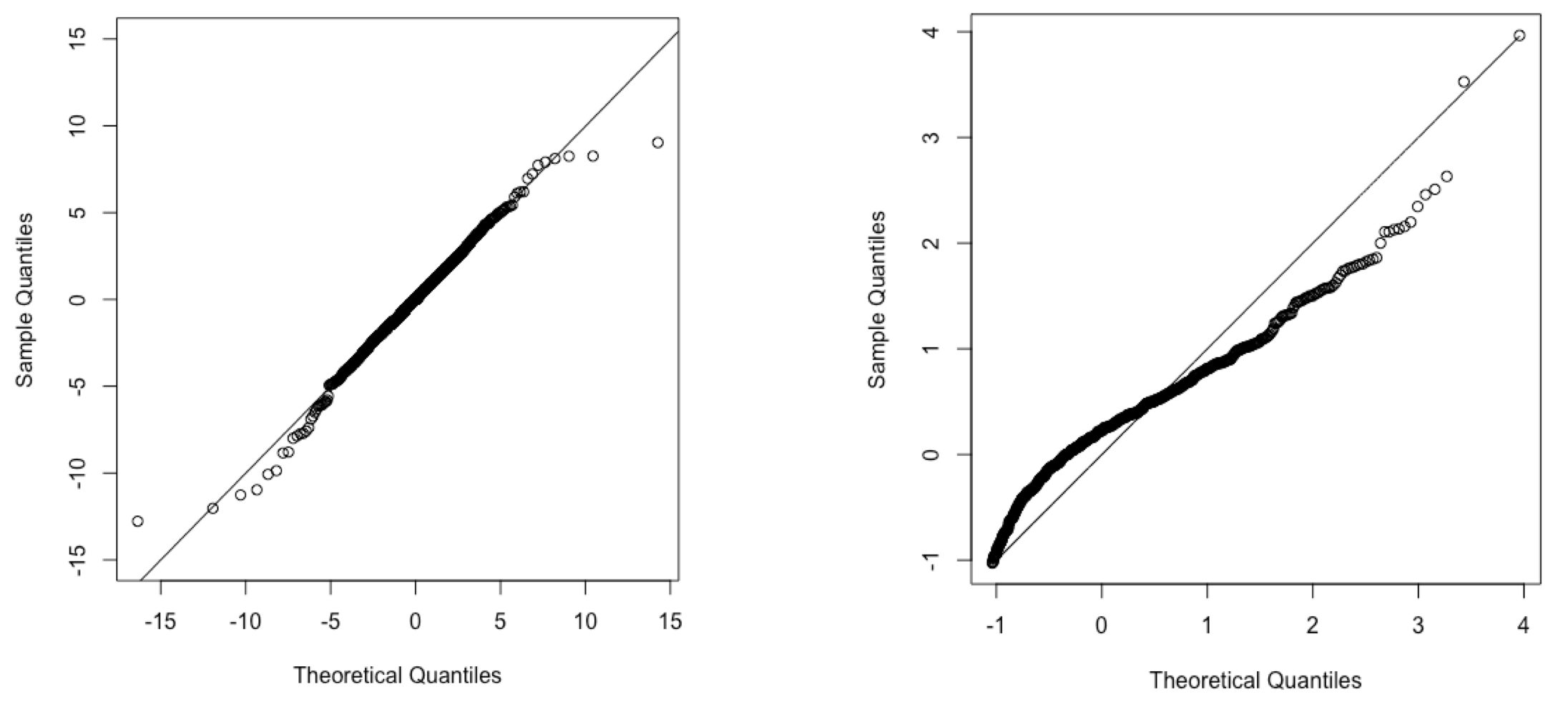}
	\centering
	\caption{QQ plots of GARCH residuals. Left: normal distribution; right: GPD distribution (samples
beyond upper 5\% threshold, g). \protect \includegraphics[height=0.5cm]{qletlogo_tr.png} {\color{blue}\href{https://github.com/QuantLet/MLvsGARCH/tree/master/MLvsGARCHecon}{MLvsGARCHecon}}}
	\label{fig:qq_plot_norm_gpd}
\end{figure}

We calibrate the parameters of $\operatorname{ARMA-GARCH}$ and $\operatorname{ARMA-EVTGARCH}$ simultaneously via quasi-maximum likelihood estimation (QMLE) \cite{Mcneil:2000} using a rolling window of fixed length of four months (2880 observations). The size of the rolling window is relatively large for  $\operatorname{GARCH}$ modelling, nevertheless, in this paper, we are using intraday data which justifies that choice. As a comparison \citet{Taylor:2016} used a rolling window of 2500 observations. The calibration of the GPD is based on the upper 5\% of the residuals of the $\operatorname{ARMA-GARCH}$, we denote the upper 5\% threshold as $g$. For illustration, residual QQ-plots based on hourly BTC loss data are given in Figure \ref{fig:qq_plot_norm_gpd}.

The estimated probability class\index{probability class}\index{exceedance probability} of for the category 2 of $Y_t$, that is $\operatorname{P}(- r_t > \operatorname{TVaR}_{t}^{\alpha})$,  is then given by $p_t = 1-F((\operatorname{TVaR}_t^{\alpha}−\hat{\mu}_{t+1})/\hat{\sigma}_{t+1})$ where $\hat{\mu}_{t+1}$ and $\hat{\sigma}_{t+1}$ are the models' mean and volatility forecasts and $F$ is the normal distribution function for the $\operatorname{ARMA-GARCH}$ and GPD for the $\operatorname{ARMA-EVTGARCH}$ model. For computing $p_t$ of $\operatorname{ARMA-EVTGARCH}$, the GPD is defined only for $x \geq g$, so if $(\operatorname{TVaR}_t^{\alpha}−\hat{\mu}_{t+1})/\hat{\sigma}_{t+1} < g$, $F$ is the normal distribution, since it is not necessary to use a GPD distribution for observations that do not belong to the tail. We recalibrate the parameters of the models every hour for each new observation in the test set.

\subsection{CARL-vol model}\index{CARL-vol}

The benefit of $\operatorname{ARMA-GARCH}$ type models is their effectiveness at catching stylized facts of financial time series and their simplicity, nonetheless, they do not directly model the exceedance probability $p_t$, but the total distribution of the return $r_t$. In order to use the advantages of $\operatorname{GARCH}$ models and to directly modelize the exceedance probability, we also fit a $\operatorname{CARL-vol}$ model from \citet{Taylor:2016} which specifies $p_t$ for a constant threshold $Q \in \mathbb{R}$ as:

$$p_{t}=\frac{0.5}{1+\operatorname{exp}\left(-x_{t}\right)}+0.5 \mathbf{I}(Q>0)$$

where $x_t =\phi_0 + \phi_1\sigma_{t}$ and we assume a $\operatorname{GARCH(1,1)}$ for the return that is,  $\sigma_{t}^{2} =\omega+ \beta_{1} \sigma_{t-1}^{2}+\alpha_{1} (r_{t-1} - \mu)^{2}$. $\phi_i$, $\alpha_i$, $\beta_i$ and $\mu$ are constant parameters.

We calibrate the parameters of $\operatorname{CARL-vol}$ with Maximum Likelihood Estimation using a Bernoulli distribution with density specified as $f(r_{t})=p_{t}^{\mathbf{I}(y_{t} \leqslant Q)}(1-p_{t})^{1-\mathbf{I}(y_{t}<Q)}$. For the model calibration, we use the same rolling window as for the $\operatorname{GARCH}$ based models of the previous section, where Q is set to $\operatorname{TVaR}_t$ of the last observation of the rolling window. Since $\operatorname{CARL-vol}$ assumes a fixed threshold $Q$ and that, in our case, it is time-varying, we also recalibrate the model every hour, estimating new parameters for each observation in an online manner.

\subsection{Local Parametric Approach (LPA)}\index{local parametric approach (LPA)}

For the calibration of the three previous models, we use a fixed rolling window, assuming time homogeneity on that interval, in the sense that the process $r_t$ follows the same structural equation at each time point defined by  $\operatorname{ARMA-GARCH}$, $\operatorname{ARMA-EVTGARCH}$ and $\operatorname{CARL-vol}$ models. This approach, of course, does not take into consideration structural breaks \cite{Bourietal:2019}. Ignoring breaks will increase the bias \cite{HILLEBRAND:2005}. Though, in order to relax time homogeneity, we can either use a non-parametric model which allows for time-varying parameters, as we do with the neural network models presented in section \ref{sec:ml_trader}, or we can use a Local Parametric Approach (LPA) thanks to the Local Change Point detection method suggested by \citet{Spokoiny:1998} and developed by  \citet{Haerdle:2003}, \citet{Mercurio:2004} and \citet{Spokoiny:2009a} in the context of volatility modelling.

As in \citet{Cizek:2009}, we assume $\operatorname{GARCH(1,1)}$, characterized by parameters $\boldsymbol{\theta}$, is a good candidate for the process $r_t$ on a local scale. For estimating $\boldsymbol{\theta}$, we apply the QMLE approach using a Gaussian distribution for the innovations. The LPA allows us to find the largest interval of homogeneity, $\mathcal{I}(t)$, i.e. the longest interval $\mathcal{I}$ with the right-end point $t$, where data do not contradict the assumption that $\boldsymbol{\theta}$ is constant. To find the change points, we use the test procedure proposed by \citet{Klochkov:2019} that uses the multiplier bootstrap method from \citet{Spokoiny:2015} to find the critical values for the likelihood ratio test, generating 100 bootstrap weights sequences for the test. The computational cost is high and our test set is quite large, thus, in practice, we choose a fixed decreasing size of 5 observations for the successive intervals tested and we only performed the LPA detection every 5 observations in the test set. This does not come without cost and automatically increases the bias of $\boldsymbol{\hat{\theta}}$, nevertheless, it decreases the bias in comparison of a fixed window size approach, as for the $\operatorname{GARCH}$ based models of the previous sections. We denote the model calibrated with LPA method as $\operatorname{LCP-GARCH}$.\index{LCP-GARCH}

\section{Applications}

\subsection{Meta strategy: stacking}\index{stacking}

In the previous section we presented six different classifiers to build trading signals for tail events hedging. Since the $\operatorname{GARCH}$ based approach is essentially linear, but MLP and LSTM are non-linear classifiers, we expect them to extract different features from the data. A direct question is then to know which model to use in which situation. To address that problem, we propose to combine all models output from the test set into a meta model with a stacking classifier. Stacking is an ensemble learning technique where the predictions of multiple classifiers (referred as level-0 classifiers) are used as new features to train a meta-classifier (level-1 classifier).

The meta-classifier\index{meta-classifier} is trained on the predictions made by level-0 classifiers on the out-of-sample data. That is, data which has not been used to train the level-0 classifiers is fed to them in order to get the predictions $p_{t} = (p_{t}^{j})$, where $1 \leq j \leq 6$ corresponds to the level-0 models and $t$ belongs to the test set indices. We then use the predictions to build input features and the corresponding labels $Y_{t}$ to build the training input-output pairs for the meta classifier.

Our meta-classifier should be explainable in order to directly interpret which level-one classifier contributes the most to the final meta-prediction. For its simplicity and interpretability, we use logistic regression. We have for all $t \in [T-n-1, T]$ where $n$ is the test set size:

$$s_t = \sum_{j=1}^{6} \beta^{j} P_{t}^{j} $$

where $P_{t}^j = \log \frac{p_{t}^{j}}{1-p_{t}^{j}}$ and $p_{t}^{j}$ is the output of each one-level classifier that is the estimated probability of class 2 of $Y_t$. Moreover, since we need to interpret the coefficients $\beta^{j}$, we rescale $p_{t}^{j}$ to $[0,1]$ range.

When using stacking technique, the issue is the multicollinearity of the prediction of the level-one classifiers. In particular, we can expect $\operatorname{ARMA-GARCH}$ and $\operatorname{ARMA-EVTGARCH}$ models to present multicollinearity in their output. To address that problem, we use L2 regularization to improve robustness, performing a Ridge type logistic regression \cite{Hoerl:1970}, where the final estimator for the model coefficients is the solution of:

$$\min _{\beta_{1} \cdots \beta_{6}} \sum_{i=T-n-1}^{T}\left(s_t -\sum_{j=1}^{6}  \beta^{j} p_{t}^{j}\right)^{2}+\lambda \sum_{j=1}^{6}  \beta^{j^2}$$

where $\lambda \in \mathbb{R}^+$ is the regularization hyper-parameter. In practice, we must finetune this parameter by performing, for example, cross-validation. In our setup, we did not tune it since cross-validation would be too computationally intensive and fixed it to $\lambda = 1$. We will denote that model as $\operatorname{Ensemble}$.\index{Ensemble} \newline

In the next sections, we present our results. They correspond to out-of-sample performance of the different models on the test set from 2019-01-01 to 2020-02-16. The neural network models are retrained every 824 observations corresponding to the size of the validation folds in order to reflect the validation performance, whereas the $\operatorname{ARMA-GARCH}$, $\operatorname{ARMA-EVTGARCH}$, $\operatorname{CARL-vol}$ and $\operatorname{Ensemble}$ are recalibrated every new observation and $\operatorname{LPA-GARCH}$ every 5 observations as stated in the previous section.

Every hour, each model generate the trading decision $\hat{s}_t$ based on the data available until $t$, then it is applied to the return, $r_{t+1}$, on the next period $[t, t+1]$. The current set-up does not account for market frictions, such as bid-ask spreads, and other associated costs, but we included the trading fees from Binance exchange, which are, at most, 0.1\% on each trade. Nonetheless, the cost associated to the fees should be very low since we only trade when a tail event is predicted.

\subsection{Benchmarks}

There are a few well documented tail risk protection strategies in the literature, such as the constant proportion portfolio insurance (CPPI) strategy \cite{Black:1987}, the Dynamic Proportion Portfolio Insurance (DPPI) or Time-varying Proportion Portfolio Insurance (TPPI) strategy \cite{Hamidi:2009}, \cite{Happersberger:2019} or the protective put strategy using options. In this paper, we compared our results with more recent and machine learning oriented work, in particular the Varspread and target VaR strategies. Finally, the last benchmark used is the simple buy-and-hold strategy, where an investor buys BTC at 2019-01-01 00:00:00 and sells it at the end of the test set at 2020-02-16 22:00:00.s

\subsection{Out-of-sample performance}

\subsubsection{Model risk evaluation}

In order to assess the performance of each strategy, we first look at the model risk which comes from the statistical performance of the classifier. A familiar metric is the Area Under the $\operatorname{ROC}$ Curve ($\operatorname{AUC}$), nevertheless, the classical $\operatorname{AUC}$\index{AUC}\index{ROC}, while taking into consideration class imbalance, does not include the cost associated with each class. For our trading application, as we stated in section \ref{sec:nn}, we have strong cost imbalance between the outcomes. Thus, for our evaluation, we use the risk-adjusted or cost-adjusted $\operatorname{AUC}$ by \citet{jorda:2011}.

The ROC curve is defined for all threshold $u$ as:

\begin{align*}
\operatorname{ROC} \colon [0, 1] &\to  [0, 1] \\
\operatorname{FPR}(u) & \mapsto  \operatorname{TPR}(u)
\end{align*}

We incorporate the average costs associated with false positives ($c_F= \bar{r}_0p_0 + \bar{r}_1p_1$) and true positives ($c_T = - \bar{r}_2p_2$) in order to get the risk-adjusted curve, denoted $\operatorname{AROC}$ and rescaled to $[0, 1]$, defined for all threshold $u$ as:\index{AROC}

\begin{align*}
\operatorname{AROC} \colon [0, 1] &\to  [0, 1] \\
c_F\operatorname{FPR}(u) & \mapsto  c_T\operatorname{TPR}(u)
\end{align*}

The risk-adjusted $\operatorname{AUC}$, denoted $\operatorname{AAUC}$\index{AAUC} is then  $\operatorname{AAUC}=\int_{0}^{1} \operatorname{AROC}(u) d u$.

In Table \ref{table:risk_auc} we show the risk-adjusted $\operatorname{AUC}$ score on the whole test period. We can see that the MLP and LSTM models have similar performance, we can make the same observation for the $\operatorname{ARMA-GARCH}$, $\operatorname{ARMA-EVTGARCH}$ and $\operatorname{LPA-GARCH}$ approach model. Nonetheless, we observe that $\operatorname{LPA-GARCH}$ performs better than $\operatorname{ARMA-GARCH}$ and $\operatorname{ARMA-EVTGARCH}$ for stable risk target where $w=2880$ or $w=4320$. This indicates that using a model with less complexity fitted on a time-homogeneous window generalized better than more complex models fitted on large window. Finally, all parametric or local parametric models perform better than the neural network based models in terms of risk-adjusted classification, which is a major drawback to the deep learning approach in our setting.

On Table \ref{table:cv_risk_auc} we present the cross-validation performance on the test set. Indeed recall, that the results from the MLP and the LSTM models are obtained by retraining the models every 824 observations in the same approach as depicted in Figure \ref{fig:time_series_cv}. For comparison, we also evaluate the other classifiers with the same sampling method. 

\begin{table}[h!]
	\begin{center}
		\begin{tabular}{ccccccccc}
			\hline
			\hline
			$\alpha$ (\%)& $w$ (hours) & $\operatorname{MLP}$ & $\operatorname{LSTM}$ & \makecell{$\operatorname{ARMA-}$\\$\operatorname{GARCH}$} & \makecell{$\operatorname{ARMA-}$\\$\operatorname{EVTGARCH}$} & \makecell{$\operatorname{LPA-}$\\$\operatorname{GARCH}$} & $\operatorname{CARL-vol}$  & Ensemble \\
			\multirow{3}{*}{1.0}  & 24 & 69 & 69 & 73 & 73 & 71 & 71 & 73 \\
			& 2880 & 62 & 67 & 69 & 72 & 72 & 69 & 74 \\
			& 4320 & 70 & 69 & 70 & 72 & 72 & 74 & 74 \\
			\multirow{3}{*}{2.5}  & 24 & 66 & 69 & 70 & 70 & 69 & 69 & 71 \\
			& 2880 & 72 & 71 & 74 & 74 & 76 & 72 & 77 \\
			& 4320 & 71 & 73 & 74 & 74 & 75 & 73 & 77 \\
			\multirow{3}{*}{5.0}  & 24 & 65 & 65 & 66 & 66 & 65 & 66 & 67 \\
			 & 2880 & 71 & 70 & 74 & 74 & 75 & 73 & 76 \\
			 & 4320 & 70 & 71 & 74 & 73 & 75 & 73 & 76 \\
			\multirow{3}{*}{10} & 24 & 63 & 63 & 63 & 63 & 60 & 62 & 64 \\
			& 2880 & 67 & 66 & 70 & 70 & 71 & 71 & 72 \\
			& 4320 & 66 & 68 & 69 & 70 & 70 & 71 & 71 \\
			\hline
			\hline
		\end{tabular}	
	\end{center}
	\caption{Risk-adjusted AUC on the test set \protect \includegraphics[height=0.5cm]{qletlogo_tr.png} {\color{blue}\href{https://github.com/QuantLet/MLvsGARCH/}{MLvsGARCH}}}
	\label{table:risk_auc}
\end{table}

We can clearly see that $\operatorname{ARMA-GARCH}$ and $\operatorname{ARMA-EVTARCH}$ outperforms MLP and LSTM classifiers for small significance level and varying risk target ($\alpha = 0.01$ or $\alpha = 0.025$ and $w = 24$) both on average and in terms of robustness. This shows that, in our setup, $\operatorname{GARCH}$ based models still outperforms non-parametric models at catching volatility clusters. On the other hand, for larger significance level and stable risk target ($\alpha = 0.05$ or $\alpha = 0.1$ and $w = 2280$ or $w=4320$), the non-parametric approaches do at least as good as the simple $\operatorname{GARCH}$ based models on average and in terms of robustness which indicates a higher generalization power.

\begin{table}[h!]
	\begin{center}
		\begin{tabular}{ccccccccc}
			\hline
			\hline
			$\alpha$ (\%)& $w$ (hours) & $\operatorname{MLP}$ & $\operatorname{LSTM}$ & \makecell{$\operatorname{ARMA-}$\\$\operatorname{GARCH}$} & \makecell{$\operatorname{ARMA-}$\\$\operatorname{EVTGARCH}$} & \makecell{$\operatorname{LPC-}$\\$\operatorname{GARCH}$} & $\operatorname{CARL-vol}$  & Ensemble \\
			
			\multirow{3}{*}{1.0} & 24 & 70 (0.3) & 70 (0.4) & 74 (0.1) & 74 (0.1) & 72 (0.2) & 72 (0.2) & 73 (0.2) \\
			 & 2880 & 58 (3.3) & 64 (3.9) & 60 (0.9) & 61 (1.3) & 61 (1.4) & 55 (1.0) & 62 (1.8) \\
			 & 4320 & 63 (3.7) & 63 (3.1) & 58 (1.2) & 62 (1.0) & 59 (1.3) & 57 (0.8) & 60 (1.6) \\
			\multirow{3}{*}{2.5} & 24 & 68 (0.4) & 70 (0.1) & 71 (0.1) & 71 (0.1) & 69 (0.1) & 70 (0.2) & 71 (0.1) \\
			 & 2880 & 66 (0.9) & 65 (1.3) & 62 (1.3) & 62 (1.2) & 64 (1.2) & 57 (1.3) & 66 (1.1) \\
			 & 4320 & 63 (0.6) & 65 (2.0) & 62 (1.4) & 62 (1.6) & 61 (1.2) & 57 (1.3) & 65 (1.6) \\
			\multirow{3}{*}{5.0} & 24 & 67 (0.4) & 67 (0.2) & 67 (0.1) & 67 (0.1) & 65 (0.2) & 66 (0.2) & 68 (0.2) \\
			 & 2880 & 65 (0.7) & 64 (0.6) & 62 (0.8) & 62 (1.0) & 63 (0.8) & 61 (1.1) & 64 (1.0) \\
			 & 4320 & 66 (0.5) & 64 (0.7) & 62 (0.9) & 61 (1.0) & 63 (0.9) & 60 (0.8) & 64 (1.0) \\
			\multirow{3}{*}{10} & 24 & 64 (0.2) & 64 (0.2) & 64 (0.1) & 64 (0.1) & 61 (0.1) & 63 (0.2) & 65 (0.2) \\
			 & 2880 & 61 (0.7) & 63 (0.8) & 61 (0.8) & 61 (0.8) & 61 (0.7) & 60 (0.7) & 62 (0.8) \\
			 & 4320 & 62 (0.7) & 64 (0.7) & 61 (0.7) & 61 (0.7) & 61 (0.8) & 60 (0.6) & 62 (0.9) \\
			\hline
			\hline
		\end{tabular}	
	\end{center}
	\caption{Average risk-adjusted AUC and (variance risk-adjusted AUC) over the testing periods \protect \includegraphics[height=0.5cm]{qletlogo_tr.png} {\color{blue}\href{https://github.com/QuantLet/MLvsGARCH/}{MLvsGARCH}}}
	\label{table:cv_risk_auc}
\end{table}

From the two tables above, we can also conclude that we benefit from combining those classifiers into a stacking classifier, since the $\operatorname{AAUC}$ of the Ensemble classifier is always higher or equal to the maximum $\operatorname{AAUC}$ of the weak classifiers on the whole test period. Moreover, in terms of cross-validation performance, the Ensemble classifier always outperforms the average performance, except for the strategy with $\operatorname{TVAR}^{0.01, 4320}$ risk target. In particular, for larger risk level ($\alpha = 0.025$, $\alpha = 0.5$ and $\alpha = 0.1$) and small window ($w=24$), the Ensemble classifier outperforms the best performing weak classifier. This result is a clear indicator that each classifier can extract information which is uncorrelated with some of the other classifiers for highly adaptive risk target. This is a sign that each classifier might be more adapted to specific market regimes, but that the Ensemble model is more general.

Overall, from those metrics, we can say that the non-parametric models present more risk than the parametric ones or the Ensemble classifier. Now, in the next section, we will relate the model risk to the economic risk via backtest performance of the associated trading strategies.

\subsubsection{Backtest performance}

\paragraph{Risk profile}
First, we must verify that the strategy associated with each classifier is verifying the target risk constraint. For this we compare, in Table \ref{table:exceedance}, the exceedance of each strategy with the risk level $\alpha$, if the exceedance is larger than $\alpha$, then the strategy does not respect the investor's risk preferences.\index{exceedance}

\begin{table}[h!]
	\begin{center}
		\hspace*{-2.5cm}
		{\footnotesize
			\begin{tabular}{ccccccccccccc}
				\hline
				\hline
				$\alpha$ (\%)& $w$ (hours) & \makecell{Buy\&\\Hold} & $\operatorname{MLP}$ & $\operatorname{LSTM}$ & \makecell{$\operatorname{ARMA-}$\\$\operatorname{GARCH}$} & \makecell{$\operatorname{ARMA-}$\\$\operatorname{EVTGARCH}$} & \makecell{$\operatorname{LPC-}$\\$\operatorname{GARCH}$} & $\operatorname{CARL-vol}$  & Ensemble & \makecell{target\\VaR norm} & \makecell{target\\VaR evt} & Varspread \\
				\hline	
				\multirow{3}{*}{1.0} & 24 & 4.3 & 0.3 & 0.4 & 0.5 & 0.4 & 0.9 & 1.2 & 1.0 & 1.4 & 1.1 & 3.6 \\
				& 2880 & 1.0 & 0.6 & 0.7 & 0.8 & 0.6 & 0.9 & 0.4 & 0.7 & 0.5 & 0.5 & 0.7 \\
				& 4320 & 0.9 & 0.7 & 0.5 & 0.7 & 0.5 & 0.8 & 0.4 & 0.7 & 0.4 & 0.4 & 0.7 \\
				\multirow{3}{*}{2.5} & 24 & 5.2 & 1.5 & 1.8 & 1.2 & 1.0 & 2.1 & 2.0 & 1.8 & 1.8 & 1.9 & 4.3 \\
				& 2880 & 2.5 & 2.1 & 1.6 & 1.6 & 1.3 & 1.2 & 1.2 & 1.7 & 1.2 & 1.3 & 2.1 \\
				& 4320 & 2.2 & 2.0 & 1.3 & 0.9 & 0.8 & 0.8 & 1.1 & 1.7 & 1.0 & 1.0 & 2.0 \\
				\multirow{3}{*}{5.0} & 24 & 7.9 & 3.6 & 3.6 & 2.8 & 3.1 & 4.5 & 3.4 & 2.5 & 2.6 & 3.6 & 5.9 \\
				& 2880 & 4.9 & 4.0 & 3.2 & 2.3 & 2.4 & 2.3 & 2.5 & 3.5 & 2.3 & 2.8 & 3.4 \\
				& 4320 & 4.5 & 3.8 & 3.1 & 1.4 & 1.5 & 3.6 & 2.3 & 3.3 & 2.2 & 2.5 & 3.3 \\
				\multirow{3}{*}{10} & 24 & 12.4 & 8.4 & 8.1 & 4.0 & 4.0 & 8.3 & 6.6 & 8.8 & 4.4 & 7.2 & 9.7 \\
				& 2880 & 9.6 & 7.9 & 7.2 & 5.6 & 5.6 & 5.1 & 5.7 & 7.5 & 4.4 & 6.1 & 7.2 \\
				& 4320 & 9.4 & 7.4 & 7.0 & 3.3 & 3.2 & 4.4 & 6.0 & 7.1 & 4.5 & 5.7 & 7.2 \\
				\hline
				\hline
			\end{tabular}	
		}
	\end{center}
	\caption{Exceedances of tail risk protection strategies in \% \protect \includegraphics[height=0.5cm]{qletlogo_tr.png} {\color{blue}\href{https://github.com/QuantLet/MLvsGARCH/}{MLvsGARCH}}}
	\label{table:exceedance}
\end{table}

All classifiers achieve the risk targets, but both benchmark target $\operatorname{VaR}$ strategies, where $\operatorname{VaR}$ is estimated with $\operatorname{ARMA-GARCH}$ and $\operatorname{ARMA-EVTGARCH}$, do not respect the constraint for  $\operatorname{TVaR}^{1\%, 24}$. The Varspread strategy is not a target VaR strategy, but we also look at its exceedance and we see that it also fails for the same risk target, $\operatorname{TVaR}^{2.5\%, 24}$ and $\operatorname{TVaR}_{5\%, 24}$, indicating difficulties for those strategies to adapt to varying risk preferences.

Now, on Table \ref{table:avg_return}, we compare the average return of the strategies associated with each classifier and the benchmarks on the whole test period. Achieving a larger average return, while having lower exceedances indicates a better risk-profile for a tail risk averse investor. First, the Ensemble strategy always have a larger average return than the average performance of the weak strategies, that is the strategies associated with the weak classifiers, except for the $\operatorname{TVaR}^{1\%, 2880}$ risk target, and has larger average return than the maximum of the weak strategies for the $\operatorname{TVaR}^{1\%, 24}$ and $\operatorname{TVaR}^{1\%, 4320}$. In general, the Ensemble strategy also outperforms the benchmark strategies, except for the target $\operatorname{VaR}$ strategies with $\operatorname{TVaR}^{1\%, 2880}$,  $\operatorname{TVaR}^{2.5\%, 24}$, $\operatorname{TVaR}^{5\%, 24}$ and $\operatorname{TVaR}^{10\%, 24}$. Only these strategies are better suited than our offered Ensemble strategy for their respective investor's profile, as they offer larger average return, while respecting the corresponding risk target. This result shows that traders can benefit from combining classical econometrics and machine learning oriented approaches in order to offer better tail risk protection strategies to their investors.

\begin{table}[h!]
	\begin{center}
		\hspace*{-2cm}
		{\footnotesize
			\begin{tabular}{ccrrrrrrrrrr}
				\hline
				\hline
				
				$\alpha$ (\%)& $w$ (hours) & $\operatorname{MLP}$ & $\operatorname{LSTM}$ & \makecell{$\operatorname{ARMA-}$\\$\operatorname{GARCH}$} & \makecell{$\operatorname{ARMA-}$\\$\operatorname{EVTGARCH}$} & \makecell{$\operatorname{LPC-}$\\$\operatorname{GARCH}$} & $\operatorname{CARL-vol}$  & Ensemble & \makecell{target\\VaR norm} & \makecell{target\\VaR evt} & Varspread \\
				
				\hline	
				\multirow{3}{*}{1.0} & 24 & 4.9 & 5.1 & 2.3 & 1.6 & 6.5 & -15.0 & 8.9 & 6.5 & 6.3 & 4.0 \\
				& 2880 & 8.0 & 5.5 & 10.0 & 4.8 & 8.6 & 17.0 & 4.4 & 7.1 & 6.3 & 4.0 \\
				& 4320 & 8.1 & 11.0 & 8.8 & 7.0 & 9.5 & 12.0 & 14.0 & 8.2 & 7.1 & 4.0 \\
				\multirow{3}{*}{2.5}  & 24 & 7.3 & 7.1 & 3.4 & 6.3 & 8.9 & -0.7 & 6.8 & 7.0 & 7.5 & 3.7 \\
				& 2880 & 6.9 & 9.9 & 3.6 & 1.4 & 2.2 & 12.0 & 8.5 & 6.4 & 6.3 & 3.7 \\
				& 4320  & 5.8 & 10.0 & 5.3 & 4.5 & 6.3 & 12.0 & 8.3 & 7.4 & 7.5 & 3.7 \\
				\multirow{3}{*}{5.0} & 24 & 7.2 & 9.2 & 5.8 & 6.7 & 9.3 & 2.9 & 7.1 & 7.8 & 8.3 & 2.7 \\
				& 2880 & 6.8 & 13.0 & 0.8 & 1.4 & 2.7 & 13.0 & 8.7 & 5.9 & 6.2 & 2.7 \\
				& 4320 & 8.5 & 12.0 & 1.5 & 2.1 & 5.1 & 16.0 & 11.0 & 6.5 & 7.2 & 2.7 \\
				\multirow{3}{*}{10} & 24 & 5.7 & 9.6 & 5.5 & 6.1 & 2.8 & 6.9 & 6.9 & 5.9 & 7.8 & 2.4 \\
				& 2880 & 10.0 & 7.5 & 1.6 & 1.8 & 5.5 & 14.0 & 6.8 & 5.5 & 6.3 & 2.4 \\
				& 4320 & 14.0 & 9.8 & 3.8 & 4.1 & 5.5 & 12.0 & 8.1 & 5.8 & 7.3 & 2.4 \\
				\hline
				\hline
			\end{tabular}
		}
	\end{center}
	\caption{Average return in \% (e-3). Buy\&Hold: 10e-3 \% \protect \includegraphics[height=0.5cm]{qletlogo_tr.png} {\color{blue}\href{https://github.com/QuantLet/MLvsGARCH/}{MLvsGARCH}}}
	\label{table:avg_return}
\end{table}

\begin{figure}[h!]
	\hspace*{-1cm}
	\includegraphics[scale=0.35]{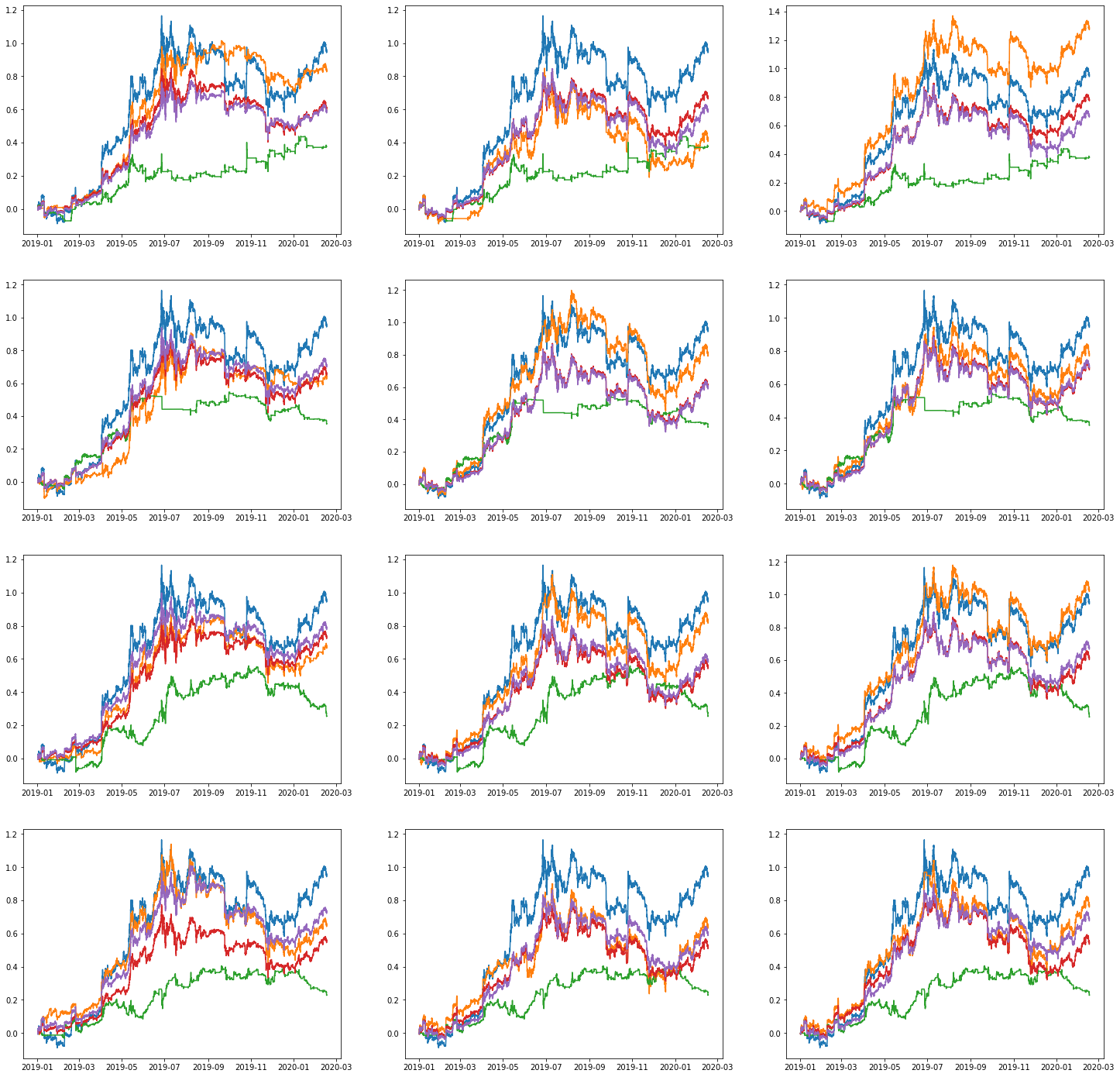}
	\centering
	\caption{\textcolor{orange}{Ensemble strategy} returns on test period for $\operatorname{TVAR}^{\alpha, w}$ with $\alpha \in \{1\%, 2.5\%, 5\%, 10\%\}$ from top to bottom and $w \in \{24, 2880, 4320\}$ from left to right with BTC \textcolor{blue}{Buy\&Hold}, \textcolor{green}{Varspread}, \textcolor{red}{target VaR norm} and \textcolor{violet}{target VaR evt} benchmarks \protect \includegraphics[height=0.5cm]{qletlogo_tr.png} {\color{blue}\href{https://github.com/QuantLet/MLvsGARCH/}{MLvsGARCH}}}
	\label{fig:strat_perf}
\end{figure}

\begin{figure}[h!]
	
	\hspace*{-1cm}
	\includegraphics[scale=0.35]{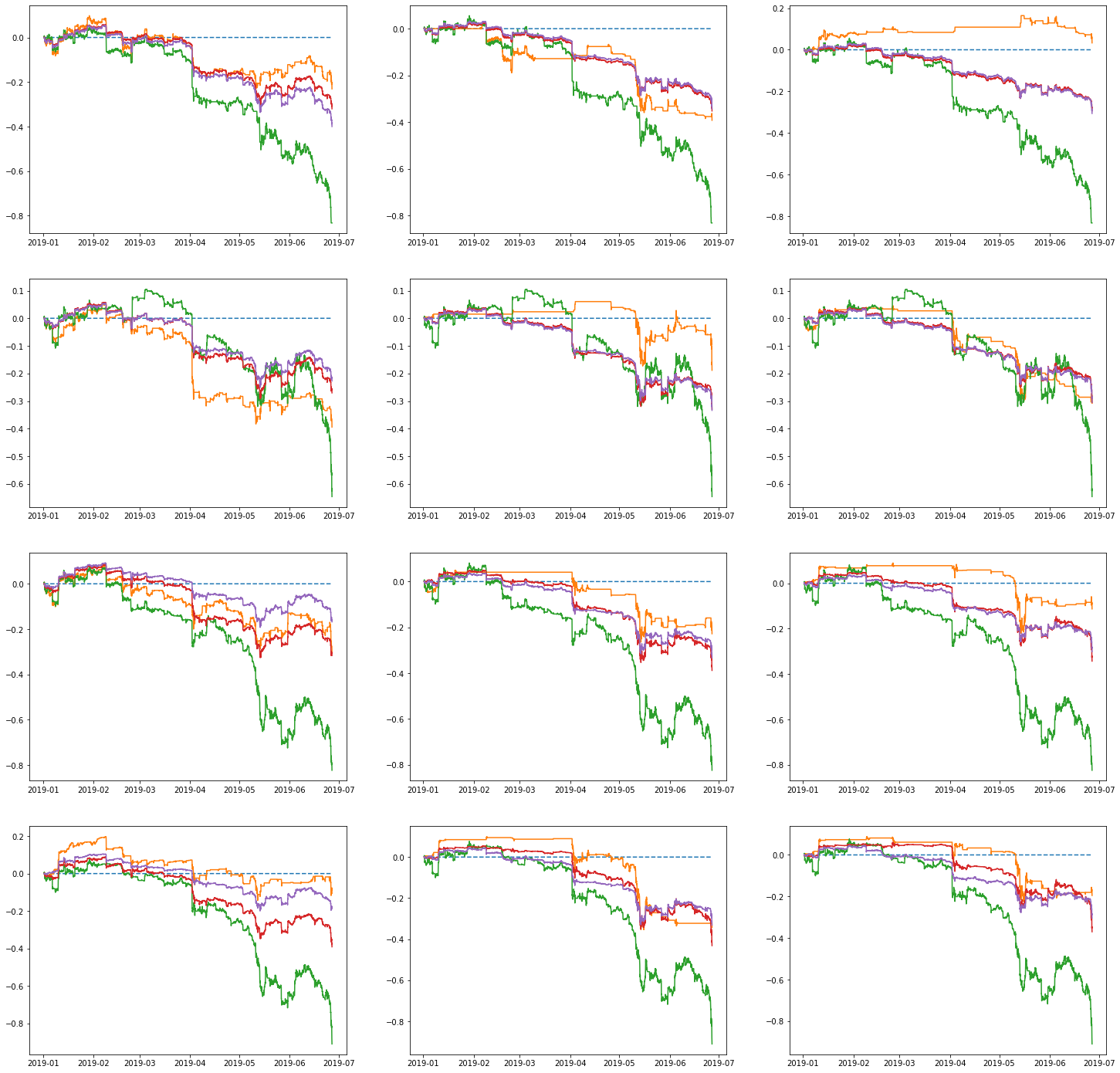}
	\centering
	\caption{\textcolor{orange}{Ensemble strategy} excess returns on uptrend period for $\operatorname{TVAR}^{\alpha, w}$ with $\alpha \in \{1\%, 2.5\%, 5\%, 10\%\}$ from top to bottom and $w \in \{24, 2880, 4320\}$ from left to right with \textcolor{green}{Varspread}, \textcolor{red}{target VaR norm} and \textcolor{violet}{target VaR evt} benchmarks \protect \includegraphics[height=0.5cm]{qletlogo_tr.png} {\color{blue}\href{https://github.com/QuantLet/MLvsGARCH/}{MLvsGARCH}}}
	\label{fig:up_excess_alpha}
\end{figure}

\paragraph{Historical performance}
On Figure \ref{fig:strat_perf}, we show the historical performance on the test period and on Figure \ref{fig:up_excess_alpha}, the excess return over the BTC Buy\&Hold benchmark in strong uptrend period. We can clearly see how Varspread strategy is too conservative in period of uptrend which severely affects its total return. While the Ensemble strategy significantly outperforms the other actively managed tail risk protection strategies in terms of total return for $\operatorname{TVAR}^{1\%, 24}$, $\operatorname{TVAR}^{1\%,4320}$, $\operatorname{TVAR}^{2.5\%,2880}$, $\operatorname{TVAR}^{2.5\%,4320}$, $\operatorname{TVAR}^{5\%,2880}$ and $\operatorname{TVAR}^{5\%,4320}$. For the other risk targets, the historical performance is similar. In the meantime, only two strategies outperforms the simple Buy\&Hold strategy on the whole test period, for  $\operatorname{TVAR}^{2.5\%,4320}$ and  $\operatorname{TVAR}^{5\%,4320}$ risk targets, managed by our proposed Ensemble model. This shows that for BTC asset, it is very hard to find the right trade-off between investment during strong uptrend and deinvestment in period of downtrend and that strategies based on simpler models, as Varspread or target VaR strategies, cannot beat the Buy\&Hold benchmark in terms of total return, as they overshoots risk in strong uptrend periods.

Indeed, as you can see on Figure \ref{fig:up_excess_alpha}, in strong uptrend period, all strategies underperforms the Buy\%Hold benchmark except for the Ensemble strategy with $\operatorname{TVAR}^{1\%, 4320}$ risk target.

\begin{figure}[h!]
	
	\hspace*{-1cm}
	\includegraphics[scale=0.35]{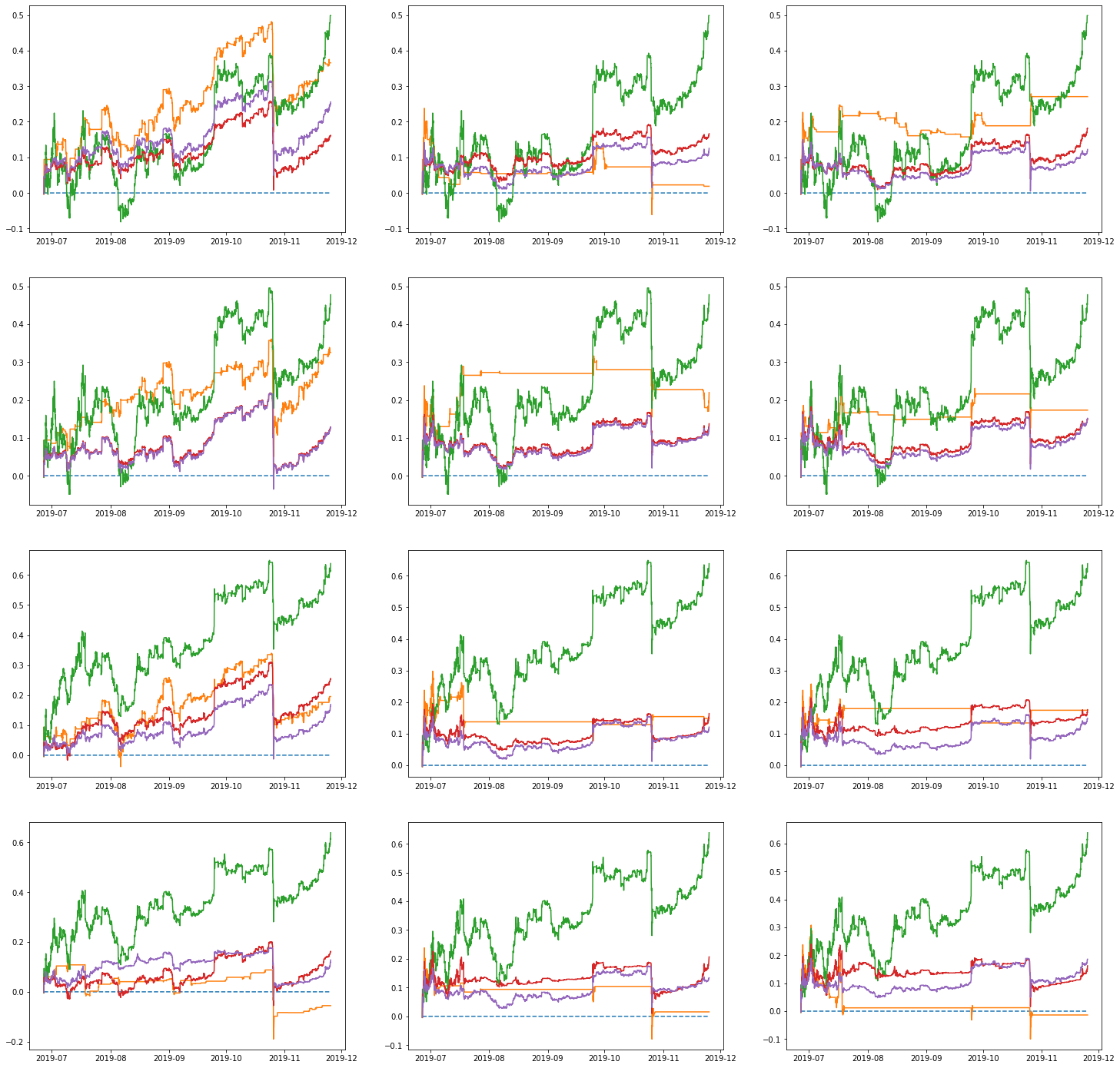}
	\centering
	\caption{\textcolor{orange}{Ensemble strategy} excess returns on downtrend period for $\operatorname{TVAR}^{\alpha, w}$ with $\alpha \in \{1\%, 2.5\%, 5\%, 10\%\}$ from top to bottom and $w \in \{24, 2880, 4320\}$ from left to right with \textcolor{green}{Varspread}, \textcolor{red}{target VaR norm} and \textcolor{violet}{target VaR evt} benchmarks \protect \includegraphics[height=0.5cm]{qletlogo_tr.png} {\color{blue}\href{https://github.com/QuantLet/MLvsGARCH/}{MLvsGARCH}}}
	\label{fig:down_excess_alpha}
\end{figure}

Nevertheless, as we can see on Figure \ref{fig:down_excess_alpha}, that, in period of downtrend from 2019-06-26 to 2019-11-25 where BTC suffered a 60\% drawdown, the Varspread strategy is very efficient at catching tail events achieving 50\%, 48\%, 64\% and 64\% excess return for $\alpha \in \{1\%, 2.5\%, 5\%, 10\%\}$ respectively even generating a positive total return of 4\% for $\alpha \in \{5\%, 10\%\}$.

\paragraph{Switching strategy}\index{switching strategy}

In order to benefit from the effective tail risk protection from the Varsrpead strategy in bear market and the well calibrated risk protection of the Ensemble strategy in bull market, we build a final switching strategy which either invest in the Varspread strategy or in the Ensemble strategy depending on the market regime. In order to decide when to switch strategy, we build a simple estimator of the trend by using a simple moving average of BTC price, defined for $n \in \mathbb{N}$ as $\mathrm{MA}_{t-1, n} = \frac{1}{n} \sum_{i=1}^{n}P_{t-1}$. We define the following trend indicator $\delta_{t}$ as in \citet{rick:2019}:

$$ \delta_{t}=\left\{\begin{array}{ll}0, & \text { if } P_{t-1} \leqslant \mathrm{MA}_{t-1, n} \\ 1, & \text { if } P_{t-1}>\mathrm{MA}_{t-1, n}\end{array}\right.$$

\begin{figure}[h!]
	\hspace*{-1cm}
	\includegraphics[scale=0.35]{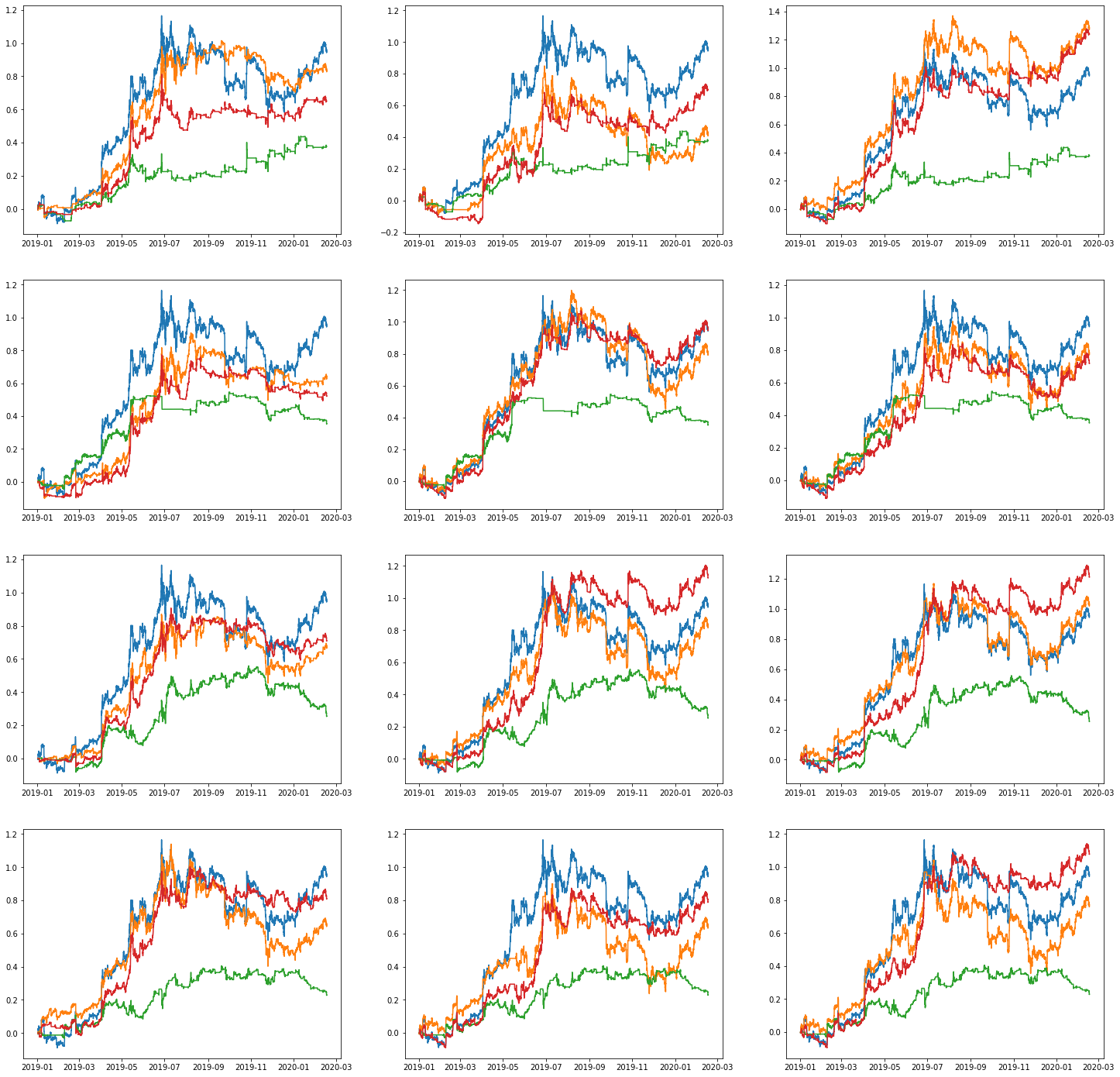}
	\centering
	\caption{Historical performance on the test period of \textcolor{orange}{Ensemble}, \textcolor{green}{Varspread}, \textcolor{red}{Switch} and \textcolor{blue}{Buy\&Hold} strategies \protect \includegraphics[height=0.5cm]{qletlogo_tr.png} {\color{blue}\href{https://github.com/QuantLet/MLvsGARCH/}{MLvsGARCH}}}
	\label{fig:strat_comb}
\end{figure}

Then we define the weights for each strategy as $w_{t}^{Ens}=\delta_{t}$ and $w_{t}^{Varspread}=1 - \delta_{t}$. The historical performance of the final meta-strategy, denoted Switch strategy, is presented on Figure \ref{fig:strat_comb}.

In Table \ref{table:all_stat}, we present different backtest statistics for the benchmark, Ensemble and final Switch strategies. First, we see that the Switch strategy has the highest total return on the testing period for relatively high risk target, but also for $\operatorname{TVAR}^{1\%, 2880}$ and $\operatorname{TVAR}^{2.5\%, 2880}$. Moreover, we can see that the Switch strategy provides an enhanced risk-return profile indicated by larger Sharpe ratio for all risk targets, than the average between the Varspread and the Ensemble strategies. For example, with $\operatorname{TVaR}^{5\%, 2880}$, the Switch strategy increased the Sharpe Ratio by 62\% and 219\% for the Ensemble and Vaspread strategies respectively. On top, the switch strategy has the best risk-return profile for all risk targets, clearly outperforming the Buy\&Hold and other benchmarks, except $\operatorname{TVAR}^{2.5\%, 24}$ where the Varspread strategy has the largest Sharpe ratio. This indicates that the switch strategy is also better suited for Markowitz type optimisation where the objective is the mean-variance performance. For the Ensemble strategy itself the results are mixed with respect to mean-variance performance, since it has a relatively high volatility for all risk targets. Indeed, finding the right trade-off between participation in bull market regimes and hedging tail events comes with a cost in terms of volatility. Since in our case, we know that the Switch and Ensemble strategies are respecting the tail risk constraint, we are interested in the Sortino ratio, which penalizes only volatility associated with negative returns and is defined as $S = \frac{R}{\sigma^-}$, where $R$ is the expected return and $\sigma^-$ is the portfolio downside standard deviation. We can see that the Switch strategy clearly outperforms other benchmarks, except for conservative strategies with $\operatorname{TVAR}^{1\%, 24}$, $\operatorname{TVAR}^{1\%, 2880}$, $\operatorname{TVAR}^{2.5\%, 24}$ and $\operatorname{TVAR}^{5\%, 24}$.

\begin{table}[h!]
	\begin{center}
		\hspace*{-2cm}
		{\footnotesize
			\begin{tabular}{cclrcccc}
				\hline
				\hline
				$\alpha$ (\%) & $w$ (hours) & Strategy & Total return & Sharpe ratio & Sortino ratio & MDD & volatility \\
				\hline
				&  & \textbf{Buy\&Hold} & \textbf{86} & \textbf{1.37} & \textbf{2.62} & \textbf{52} & \textbf{0.69} \\
				\hline
				\multirow{15}{*}{1.0} & \multirow{5}{*}{24} & Ensemble & \textbf{\color{green}76} & 1.52 & \textbf{\color{green}2.52} & 31 & \textbf{\color{red}0.54} \\
				&  & Switch & 59 & \textbf{\color{green}1.57} & 2.29 & 30 & 0.41 \\
				&  & target VaR evt & 54 & 1.28 & 2.13 & 33 & 0.46 \\
				&  & target VaR norm & 55 & 1.24 & 2.10 & \textbf{\color{red}40} & 0.49 \\
				&  & Varspread & \textbf{\color{red}34} & \textbf{\color{red}1.14} & \textbf{\color{red}1.53} & \textbf{\color{green}17} & \textbf{\color{green}0.33} \\
				\cline{2-8}
				& \multirow{5}{*}{2880} & Ensemble & 38 & \textbf{\color{red}0.70} & \textbf{\color{red}1.44} & \textbf{\color{red}53} & \textbf{\color{red}0.60} \\
				&  & Switch & \textbf{\color{green}63}  & \textbf{\color{green}1.50} & 2.36 & 27 & 0.46 \\
				&  & target VaR evt & 54 & 1.16 & 2.15 & 45 & 0.50 \\
				&  & target VaR norm & 61 & 1.32 & \textbf{\color{green}2.37} & 40 & 0.50 \\
				&  & Varspread & \textbf{\color{red}34} & 1.14 & 1.53 & \textbf{\color{green}17} &  \textbf{\color{green}0.33} \\
				\cline{2-8}
				& \multirow{5}{*}{4320} & Ensemble & \textbf{\color{green}115} & 1.97 & 3.91 & \textbf{\color{red}43} & \textbf{\color{red}0.64} \\
				&  & Switch & 111 & \textbf{\color{green}2.39} & \textbf{\color{green}3.95} & 23 & 0.51 \\
				&  & target VaR evt & 60 & 1.31 & 2.37 & \textbf{\color{red}43} & 0.50 \\
				&  & target VaR norm & 71 & 1.54 & 2.75 & 39 & 0.50 \\
				&  & Varspread & \textbf{\color{red}34} & \textbf{\color{red}1.14} & \textbf{\color{red}1.53} & \textbf{\color{green}17} &  \textbf{\color{green}0.33} \\
				\hline
				\multirow{15}{*}{2.5} & \multirow{5}{*}{24} & Ensemble & 58 & \textbf{\color{red}1.14} & \textbf{\color{red}1.71} & 36 & \textbf{\color{red}0.55} \\
				&  & Switch & 47 & 1.27 & 1.90 & 27 & 0.41 \\
				&  & target VaR evt & \textbf{\color{green}64} & \textbf{\color{green}1.40} & \textbf{\color{green}2.37} & \textbf{\color{red}42} & 0.50 \\
				&  & target VaR norm & 60 & 1.32 & 2.21 & \textbf{\color{red}42} & 0.50 \\
				&  & Varspread & \textbf{\color{red}34} & 1.35 & 2.23 & \textbf{\color{green}18} & \textbf{\color{green}0.26} \\
				\cline{2-8}
				& \multirow{5}{*}{2880} & Ensemble & 73 & 1.34 & 2.71 & \textbf{\color{red}55} & \textbf{\color{red}0.59} \\
				&  & Switch & \textbf{\color{green}85} & \textbf{\color{green}2.05}& \textbf{\color{green}4.15} & 32 & 0.46 \\
				&  & target VaR evt & 54 & \textbf{\color{red}1.19} & \textbf{\color{red}2.21} & 45 & 0.50 \\
				&  & target VaR norm & 55 & 1.23 & 2.30 & 45 & 0.49 \\
				&  & Varspread & \textbf{\color{red}34} & 1.35 & 2.23 & \textbf{\color{green}18} & \textbf{\color{green}0.26} \\
				\cline{2-8}
				& \multirow{5}{*}{4320} & Ensemble & \textbf{\color{green}71} & \textbf{\color{red}1.26} & 2.54 & \textbf{\color{red}45} & \textbf{\color{red}0.62} \\
				&  & Switch & 65 & \textbf{\color{green}1.50} & \textbf{\color{green}2.91} & 30 & 0.47 \\
				&  & target VaR evt & 64 & 1.43 & 2.55 & 41 & 0.49 \\
				&  & target VaR norm & 63 & 1.42 & 2.55 & 41 & 0.49 \\
				&  & Varspread & \textbf{\color{red}32} & 1.35 & \textbf{\color{red}2.23} & \textbf{\color{green}18} & \textbf{\color{green}0.26} \\
				\hline
				\multirow{15}{*}{5.0} & \multirow{5}{*}{24} & Ensemble & 61 & 1.25 & 1.86 & \textbf{\color{red}39} & \textbf{\color{red}0.53} \\
				&  & Switch & 64 & \textbf{\color{green}1.64} & 2.69 & 29 & 0.43 \\
				&  & target VaR evt & \textbf{\color{green}71} & 1.58 & \textbf{\color{green}2.82} & 39 & 0.49 \\
				&  & target VaR norm & 66 & 1.53 & 2.67 & 34 & 0.47 \\
				&  & Varspread & \textbf{\color{red}23} & \textbf{\color{red}0.70} & \textbf{\color{red}1.33} & \textbf{\color{green}27} & \textbf{\color{green}0.36} \\
				\cline{2-8}
				& \multirow{5}{*}{2880} & Ensemble & 75 & 1.38 & 2.60 & \textbf{\color{red}54} & \textbf{\color{red}0.59} \\
				&  & Switch & \textbf{\color{green}101} & \textbf{\color{green}2.23} & \textbf{\color{green}5.08} & 27 & 0.50 \\
				&  & target VaR evt & 54 & 1.18 & 2.19 & 45 & 0.49 \\
				&  & target VaR norm & 50 & 1.17 & 2.20 & 44 & 0.47 \\
				&  & Varspread & \textbf{\color{red}23} & \textbf{\color{red}0.70} & \textbf{\color{red}1.33} & \textbf{\color{green}27} &  \textbf{\color{green}0.36} \\
				\cline{2-8}
				& \multirow{5}{*}{4320} & Ensemble & 92 & 1.68 & 2.92 & \textbf{\color{red}49} & \textbf{\color{red}0.61} \\
				&  & Switch & \textbf{\color{green}108} & \textbf{\color{green}2.38} & \textbf{\color{green}5.09} & 23 & 0.50 \\
				&  & target VaR evt & 61 & 1.38 & 2.48 & 42 & 0.49 \\
				&  & target VaR norm & 56 & 1.29 & 2.39 & 42 & 0.47 \\
				&  & Varspread & \textbf{\color{red}23} & \textbf{\color{red}0.70} & \textbf{\color{red}1.33} & \textbf{\color{green}27} &  \textbf{\color{green}0.36} \\
				\hline
				\multirow{15}{*}{10} & \multirow{5}{*}{24} & Ensemble & 59 & 1.07 & 1.94 & \textbf{\color{red}55} & \textbf{\color{red}0.61} \\
				&  & Switch & \textbf{\color{green}73} & \textbf{\color{green}1.70} & \textbf{\color{green}3.03} & 26 & 0.47 \\
				&  & target VaR evt & 66 & 1.43 & 2.65 & 42 &  0.51 \\
				&  & target VaR norm & 51 & 1.22 & 2.01 & 39 &  0.45 \\
				&  & Varspread & \textbf{\color{red}21} & \textbf{\color{red}0.74} & \textbf{\color{red}1.18} & \textbf{\color{green}17} &  \textbf{\color{green}0.30} \\
				\cline{2-8}
				& \multirow{5}{*}{2880} & Ensemble & 58 & 1.07 & 1.95 & \textbf{\color{red}52} &  \textbf{\color{red}0.60} \\
				&  & Switch & \textbf{\color{green}71} & \textbf{\color{green}1.69} & \textbf{\color{green}3.33} & 27 &  0.46 \\
				&  & target VaR evt & 54 & 1.21 & 2.29 & 44 & 0.49 \\
				&  & target VaR norm & 47 & 1.18 & 2.16 & 40 & 0.44 \\
				&  & Varspread & \textbf{\color{red}21} & \textbf{\color{red}0.74} & \textbf{\color{red}1.18} & \textbf{\color{green}17} & \textbf{\color{green}0.30} \\
				\cline{2-8}
				& \multirow{5}{*}{4320} & Ensemble & 70 & 1.32 & 2.33 & \textbf{\color{red}53} & \textbf{\color{red}0.58} \\
				&  & Switch & \textbf{\color{green}97} & \textbf{\color{green}2.27} & \textbf{\color{green}4.50} & 23 & 0.47 \\
				&  & target VaR evt & 62 & 1.41 & 2.57 & 40 & 0.48 \\
				&  & target VaR norm & 50 & 1.25 & 2.28 & 44 & 0.43 \\
				&  & Varspread & \textbf{\color{red}21} & \textbf{\color{red}0.74} & \textbf{\color{red}1.18} & \textbf{\color{green}17} & \textbf{\color{green}0.30} \\
				
				\hline
				\hline
			\end{tabular}
		}
	\end{center}
	\caption{Backtest performance of switch strategies: Total return (\%), Sharpe ratio, Sortino ratio, Max Drawdown (MDD), Valut-at-Risk and Volatility of different strategies on the test period. The Total return, Sharpe and Sortino ratio are annualized. Legend: \textbf{\color{green}winner} / \textbf{\color{red}loser} \protect \includegraphics[height=0.5cm]{qletlogo_tr.png} {\color{blue}\href{https://github.com/QuantLet/MLvsGARCH/}{MLvsGARCH}}}
	\label{table:all_stat}
\end{table}

\clearpage

\newpage

\section{Conclusion}

We build a dynamic trading strategy which goals is to protect against severe losses by targeting a maximum Value-At-Risk defined for any significance level. For this, we model distributional properties of BTC returns in time with non-parametric models such as MLP and LSTM neural networks and compare their performance to local-parametric GARCH, parametric GARCH fitted with various distribution for the innovation and CARL-Vol models. In our approach, we directly take into account varying investor's risk preferences and optimize the strategy accordingly, allowing us to compare the models for multiple risk targets.

First, we can conclude that for relatively stable risk preferences, parametric and local parametric approaches have a better generalization than non-parametric methods, but for unstable preferences they work better, taking into account time varying risk targets and adapting to sudden changes in the market. On top, we show that, for BTC returns, a general model performing well enough in all market regimes is too hard to find. A tail loss protection strategy based on a single model, while often outperforming a simple Buy\&Hold strategy, is still not optimal, either overshooting the risk target in periods of losses or being too conservative in bull market regimes.
That is why, we show the benefits of combining multiple models in an Ensemble classifier, which improves generalization with a higher prediction accuracy and economic gains with an enhanced risk-return profile. Indeed, the Ensemble based strategy combines advantages of classical GARCH tools, which catches better volatility clusters than more complex MLP or LSTM, and  Deep learning tools which can adapt to time-varying properties of the data distribution. Further, we show that tail loss protection strategy also suits Markowitz type investors when we combine conservative strategies, such as Varspread and more aggressive ones, such as Ensemble strategy, into a switching strategy.

For the purpose of tail loss protection, our proposed Ensemble strategy can be improved. Here, we measure risk with Value-At-Risk, nonetheless, other risk metrics can be considered for optimization, in particular Expected Shortfall or Maximum Drawdown. On top, it would be interesting to open the Ensemble learning tool box and to study how the different classifiers interacts which other in time, to better understand when their corresponding tail loss protection strategy might be suited for the respective market regime.

\newpage
\bibliography{bibliography}

\begin{thebibliography}{}

\bibitem[\protect\citeauthoryear{Black}{Black}{1976}]{Black:1976}
Black, F. (1976).
\newblock Studies of stock price volatility changes.
\newblock In {\em Proceedings of the 1976 Meetings of the American Statistical
  Association}, pp.\  177--189. Business and Economics Statistics Section.

\bibitem[\protect\citeauthoryear{Black and Jones}{Black and
  Jones}{1987}]{Black:1987}
Black, F. and R.~W. Jones (1987).
\newblock Simplifying portfolio insurance.
\newblock ~{\em 14\/}(1), 48--51.

\bibitem[\protect\citeauthoryear{Bollerslev}{Bollerslev}{1986}]{Bollerslev:1986}
Bollerslev, T. (1986).
\newblock Generalized autoregressive conditional heteroskedasticity.
\newblock {\em Journal of Econometrics\/}~{\em 31\/}(3), 307 -- 327.

\bibitem[\protect\citeauthoryear{Bollerslev and Todorov}{Bollerslev and
  Todorov}{2011}]{boll:2011}
Bollerslev, T. and V.~Todorov (2011).
\newblock Tails, fears, and risk premia.
\newblock {\em The Journal of Finance\/}~{\em 66\/}(6), 2165--2211.

\bibitem[\protect\citeauthoryear{Bollerslev, Todorov, and Xu}{Bollerslev
  et~al.}{2015}]{boll:2015}
Bollerslev, T., V.~Todorov, and L.~Xu (2015).
\newblock Tail risk premia and return predictability.
\newblock {\em Journal of Financial Economics\/}~{\em 118\/}(1), 113--134.

\bibitem[\protect\citeauthoryear{Bouri, Gil-Alana, Gupta, and Roubaud}{Bouri
  et~al.}{2019}]{Bourietal:2019}
Bouri, E., L.~Gil-Alana, R.~Gupta, and D.~Roubaud (2019).
\newblock Modelling long memory volatility in the bitcoin market: Evidence of
  persistence and structural breaks.
\newblock {\em International Journal of Finance \& Economics\/}~{\em 24\/}(1),
  412--426.

\bibitem[\protect\citeauthoryear{Chen, Chen, Härdle, Lee, and Ong}{Chen
  et~al.}{2018}]{Chen:2016}
Chen, S., C.~Y.-H. Chen, W.~K. Härdle, T.~Lee, and B.~Ong (2018).
\newblock Econometric analysis of a cryptocurrency index for portfolio
  investment.
\newblock In D.~{Lee Kuo Chuen} and R.~Deng (Eds.), {\em Handbook of
  Blockchain, Digital Finance, and Inclusion, Volume 1}, pp.\  175 -- 206.
  Academic Press.

\bibitem[\protect\citeauthoryear{Christie}{Christie}{1982}]{Christie:1982}
Christie, A. (1982).
\newblock The stochastic behavior of common stock variances: Value, leverage
  and interest rate effects.
\newblock {\em Journal of Financial Economics\/}~{\em 10\/}(4), 407--432.

\bibitem[\protect\citeauthoryear{Christoffersen and Diebold}{Christoffersen and
  Diebold}{2006}]{Christoffersen:2006}
Christoffersen, P.~F. and F.~X. Diebold (2006).
\newblock Financial asset returns, direction-of-change forecasting, and
  volatility dynamics.
\newblock {\em Management Science\/}~{\em 52\/}(8), 1273--1287.

\bibitem[\protect\citeauthoryear{Chung and Hong}{Chung and
  Hong}{2007}]{Chung:2007}
Chung, J. and Y.~Hong (2007).
\newblock Model-free evaluation of directional predictability in foreign
  exchange markets.
\newblock {\em Journal of Applied Econometrics\/}~{\em 22\/}(5), 855--889.

\bibitem[\protect\citeauthoryear{Cizek, H{\"a}rdle, and Spokoiny}{Cizek
  et~al.}{2009}]{Cizek:2009}
Cizek, P., W.~K. H{\"a}rdle, and V.~Spokoiny (2009).
\newblock Adaptive pointwise estimation in time-inhomogeneous conditional
  heteroscedasticity models.
\newblock {\em The Econometrics Journal\/}~{\em 12\/}(2), 248--271.

\bibitem[\protect\citeauthoryear{de~Prado}{de~Prado}{2018}]{Prado:2018}
de~Prado, M.~L. (2018).
\newblock {\em Advances in Financial Machine Learning\/} (1st ed.).
\newblock Wiley Publishing.

\bibitem[\protect\citeauthoryear{Franke}{Franke}{1999}]{Franke:1999}
Franke, J. (1999).
\newblock Nonlinear and nonparametric methods for analyzing financial time
  series.
\newblock In P.~Kall and H.-J. L{\"u}thi (Eds.), {\em Operations Research
  Proceedings 1998}, Berlin, Heidelberg, pp.\  271--282. Springer Berlin
  Heidelberg.

\bibitem[\protect\citeauthoryear{Franke, H{\"a}rdle, and Hafner}{Franke
  et~al.}{2019}]{Franke:2019}
Franke, J., W.~K. H{\"a}rdle, and C.~M. Hafner (2019).
\newblock {\em Statistics of Financial Markets - An introduction\/} (5th ed.).
\newblock Springer Verlag.

\bibitem[\protect\citeauthoryear{Hakansson}{Hakansson}{1971}]{Hakansson:1971}
Hakansson, N. (1971).
\newblock Capital growth and the mean-variance approach to portfolio selection.
\newblock {\em Journal of Financial and Quantitative Analysis\/}~{\em 6\/}(1),
  517--557.

\bibitem[\protect\citeauthoryear{Hamidi, Jurzenko, and Maillet}{Hamidi
  et~al.}{2009}]{Hamidi:2009}
Hamidi, B., E.~Jurzenko, and B.~Maillet (2009).
\newblock A caviar modelling for a simple time-varying proportion portfolio
  insurance strategy.
\newblock {\em Bankers, Markets and Investors\/}~{\em 102\/}(4–21).

\bibitem[\protect\citeauthoryear{Happersberger, Lohre, and Nolte}{Happersberger
  et~al.}{2019}]{Happersberger:2019}
Happersberger, D., H.~Lohre, and I.~Nolte (2019).
\newblock Estimating portfolio risk for tail risk protection strategies.
\newblock {\em European Financial Management\/}.

\bibitem[\protect\citeauthoryear{H{\"a}rdle, Herwatz, and Spokoiny}{H{\"a}rdle
  et~al.}{2003}]{Haerdle:2003}
H{\"a}rdle, W.~K., H.~Herwatz, and V.~Spokoiny (2003).
\newblock Time inhomogeneous multiple volatility modelling.
\newblock {\em Journal of Financial Econometrics\/}~{\em 1}, 55--99.

\bibitem[\protect\citeauthoryear{Hillebrand}{Hillebrand}{2005}]{HILLEBRAND:2005}
Hillebrand, E. (2005).
\newblock Neglecting parameter changes in garch models.
\newblock {\em Journal of Econometrics\/}~{\em 129\/}(1), 121 -- 138.
\newblock Modelling structural breaks.

\bibitem[\protect\citeauthoryear{Hochreiter and Schmidhuber}{Hochreiter and
  Schmidhuber}{1997}]{hoch:1997}
Hochreiter, S. and J.~Schmidhuber (1997, November).
\newblock Long short-term memory.
\newblock {\em Neural Comput.\/}~{\em 9\/}(8), 1735–1780.

\bibitem[\protect\citeauthoryear{Hoerl and Kennard}{Hoerl and
  Kennard}{1970}]{Hoerl:1970}
Hoerl, A.~E. and R.~W. Kennard (1970).
\newblock Ridge regression: Biased estimation for nonorthogonal problems.
\newblock {\em Technometrics\/}~{\em 12\/}(1), 55--67.

\bibitem[\protect\citeauthoryear{Honegger and Wijewickreme}{Honegger and
  Wijewickreme}{2013}]{Honegger:2013}
Honegger, D. and D.~Wijewickreme (2013).
\newblock Seismic risk assessment for oil and gas pipelines.
\newblock In S.~Tesfamariam and K.~Goda (Eds.), {\em Handbook of Seismic Risk
  Analysis and Management of Civil Infrastructure Systems}, Woodhead Publishing
  Series in Civil and Structural Engineering, pp.\  682 -- 715. Woodhead
  Publishing.

\bibitem[\protect\citeauthoryear{Jordà and Taylor}{Jordà and
  Taylor}{2011}]{jorda:2011}
Jordà, O. and A.~M. Taylor (2011, June).
\newblock Performance evaluation of zero net-investment strategies.
\newblock Working Paper 17150, National Bureau of Economic Research.

\bibitem[\protect\citeauthoryear{Kahneman and Tversky}{Kahneman and
  Tversky}{1979}]{kahn:1979}
Kahneman, D. and A.~Tversky (1979).
\newblock Prospect theory: An analysis of decision under risk.
\newblock {\em Econometrica\/}~{\em 47\/}(2), 263--91.

\bibitem[\protect\citeauthoryear{Kelly}{Kelly}{1956}]{Kelly:1956}
Kelly, J.~L. (1956).
\newblock A new interpretation of information rate.
\newblock {\em The Bell System Technical Journal\/}~{\em 35\/}(4), 917--926.

\bibitem[\protect\citeauthoryear{Kim and Won}{Kim and Won}{2018}]{Kim:2018}
Kim, H.~Y. and C.~H. Won (2018).
\newblock Forecasting the volatility of stock price index: A hybrid model
  integrating lstm with multiple garch-type models.
\newblock {\em Expert Systems with Applications\/}~{\em 103}, 25 -- 37.

\bibitem[\protect\citeauthoryear{Kingma and Ba}{Kingma and
  Ba}{2015}]{Kingma:2015}
Kingma, D.~P. and J.~Ba (2015).
\newblock Adam: A method for stochastic optimization.
\newblock {\em CoRR\/}~{\em abs/1412.6980}.

\bibitem[\protect\citeauthoryear{Klochkov, H{\"a}rdle, and Xu}{Klochkov
  et~al.}{2019}]{Klochkov:2019}
Klochkov, Y., W.~K. H{\"a}rdle, and X.~Xu (2019).
\newblock Localizing multivariate {CAV}iar.
\newblock {\em SFB 649 Discussion Paper, submitted to Journal of Applied
  Econometrics\/}~{\em 2019\/}(007).

\bibitem[\protect\citeauthoryear{Kozhan, Neuberger, and Schneider}{Kozhan
  et~al.}{2013}]{kozh:2013}
Kozhan, R., A.~Neuberger, and P.~Schneider (2013, 09).
\newblock {The Skew Risk Premium in the Equity Index Market}.
\newblock {\em The Review of Financial Studies\/}~{\em 26\/}(9), 2174--2203.

\bibitem[\protect\citeauthoryear{Krawczyk}{Krawczyk}{2016}]{Krawczyk:2016}
Krawczyk, B. (2016).
\newblock Learning from imbalanced data: open challenges and future directions.
\newblock {\em Progress in Artificial Intelligence\/}~{\em 5\/}(4), 221--232.

\bibitem[\protect\citeauthoryear{Kunreuther}{Kunreuther}{2002}]{Kunreuther:2002}
Kunreuther, H. (2002).
\newblock Risk analysis and risk management in an uncertain world.
\newblock {\em Risk Analysis\/}~{\em 22\/}(4), 655--664.

\bibitem[\protect\citeauthoryear{Lambert, Matalas, Ling, Haimes, and
  Li}{Lambert et~al.}{1994}]{Lambert:1994}
Lambert, J., N.~Matalas, C.~Ling, Y.~Haimes, and D.~Li (1994).
\newblock Selection of probability distributions in characterizing risk of
  extreme events.
\newblock {\em Risk Analysis\/}~{\em 14\/}(5), 731--742.

\bibitem[\protect\citeauthoryear{Leshno, Lin, Pinkus, and Schocken}{Leshno
  et~al.}{1993}]{Leshno:1993}
Leshno, M., V.~Y. Lin, A.~Pinkus, and S.~Schocken (1993).
\newblock Multilayer feedforward networks with a nonpolynomial activation
  function can approximate any function.
\newblock {\em Neural Networks\/}~{\em 6\/}(6), 861 -- 867.

\bibitem[\protect\citeauthoryear{Linton and Whang}{Linton and
  Whang}{2007}]{Linton:2007}
Linton, O. and Y.-J. Whang (2007).
\newblock The quantilogram: With an application to evaluating directional
  predictability.
\newblock {\em Journal of Econometrics\/}~{\em 141\/}(1), 250--282.

\bibitem[\protect\citeauthoryear{Longin and Solnik}{Longin and
  Solnik}{2001}]{longin:2001}
Longin, F. and B.~Solnik (2001).
\newblock Extreme correlation of international equity markets.
\newblock {\em The Journal of Finance\/}~{\em 56\/}(2), 649--676.

\bibitem[\protect\citeauthoryear{Markowitz}{Markowitz}{1952}]{Markowitz:1952}
Markowitz, H. (1952).
\newblock Portfolio selection.
\newblock {\em Journal of Finance\/}~{\em 7\/}(1), 77--91.

\bibitem[\protect\citeauthoryear{Mason, Galpin, Goddard, Graham, and
  Rajartnam}{Mason et~al.}{2007}]{Mason:2007}
Mason, S.~J., J.~S. Galpin, L.~Goddard, N.~E. Graham, and B.~Rajartnam (2007,
  02).
\newblock {Conditional Exceedance Probabilities}.
\newblock {\em Monthly Weather Review\/}~{\em 135\/}(2), 363--372.

\bibitem[\protect\citeauthoryear{McNeil and Frey}{McNeil and
  Frey}{2000}]{Mcneil:2000}
McNeil, A. and R.~Frey (2000).
\newblock Estimation of tail-related risk measures for heteroscedastic
  financial time series: an extreme value approach.
\newblock {\em Journal of Empirical Finance\/}~{\em 7\/}(3), 271 -- 300.
\newblock Special issue on Risk Management.

\bibitem[\protect\citeauthoryear{Mercurio and Spokoiny}{Mercurio and
  Spokoiny}{2004}]{Mercurio:2004}
Mercurio, D. and V.~Spokoiny (2004).
\newblock Statistical inference for time-inhomogeneous volatility models.
\newblock {\em The Annals of Statistics\/}~{\em 32}, 577–602.

\bibitem[\protect\citeauthoryear{Packham, Papenbrock, Schwendner, and
  Woebbeking}{Packham et~al.}{2017}]{pack:2017}
Packham, N., J.~Papenbrock, P.~Schwendner, and F.~Woebbeking (2017).
\newblock Tail-risk protection trading strategies.
\newblock {\em Quantitative Finance\/}~{\em 17\/}(5), 729--744.

\bibitem[\protect\citeauthoryear{Rickenberg}{Rickenberg}{2019}]{rick:2019}
Rickenberg, L. (2019, 01).
\newblock Tail risk targeting: Target var and cvar strategies.
\newblock {\em SSRN Electronic Journal\/}.

\bibitem[\protect\citeauthoryear{Spokoiny}{Spokoiny}{1998}]{Spokoiny:1998}
Spokoiny, V. (1998).
\newblock Estimation of a function with discontinuities via local polynomial
  fit with an adaptive window choice.
\newblock {\em The Annals of Statistics\/}~{\em 26}, 1356–78.

\bibitem[\protect\citeauthoryear{Spokoiny}{Spokoiny}{009a}]{Spokoiny:2009a}
Spokoiny, V. (2009a).
\newblock Multiscale local change point detection with applications to
  value-at-risk.
\newblock {\em The Annals of Statistics\/}~{\em 37}, 1405--1436.

\bibitem[\protect\citeauthoryear{Spokoiny and Zhilova}{Spokoiny and
  Zhilova}{2015}]{Spokoiny:2015}
Spokoiny, V. and M.~Zhilova (2015).
\newblock Bootstrap confidence sets under model misspecification.
\newblock {\em The Annals of Statistics\/}~{\em 43\/}(1), 2653–2675.

\bibitem[\protect\citeauthoryear{Srivastava, Hinton, Krizhevsky, Sutskever, and
  Salakhutdinov}{Srivastava et~al.}{2014}]{Srivastava:2014}
Srivastava, N., G.~Hinton, A.~Krizhevsky, I.~Sutskever, and R.~Salakhutdinov
  (2014).
\newblock Dropout: A simple way to prevent neural networks from overfitting.
\newblock {\em Journal of Machine Learning Research\/}~{\em 15\/}(56),
  1929--1958.

\bibitem[\protect\citeauthoryear{Taylor and Yu}{Taylor and
  Yu}{2016}]{Taylor:2016}
Taylor, J. and K.~Yu (2016).
\newblock Using auto-regressive logit models to forecast the exceedance
  probability for financial risk management.
\newblock {\em Journal of the Royal Statistical Society: Series A (Statistics
  in Society)\/}~{\em 179\/}(4), 1069--1092.

\bibitem[\protect\citeauthoryear{Wen, He, Chen, and Li}{Wen
  et~al.}{2014}]{wen:2014}
Wen, F., Z.~He, X.~Chen, and J.~Li (2014, 01).
\newblock Investors’ risk preference characteristics and conditional
  skewness.
\newblock {\em Mathematical Problems in Engineering\/}.

\bibitem[\protect\citeauthoryear{Zhang, Zohren, and Roberts}{Zhang
  et~al.}{2020}]{Zhang:2020}
Zhang, Z., S.~Zohren, and S.~Roberts (2020).
\newblock Deep learning for portfolio optimization.
\newblock {\em The Journal of Financial Data Science\/}.

\end{thebibliography}

\thispagestyle{empty}
\printindex

\end{document}